\documentclass[11pt]{article}
\pdfoutput=1 
\usepackage{dsfont}
\usepackage{comment}
\usepackage{graphicx}
\usepackage{latexsym,comment}
\usepackage[english]{babel}
 \usepackage{amssymb,dsfont,amsmath,amsfonts,hyperref}
\usepackage{slashed}
\usepackage{color} 
\usepackage{cite}
\usepackage[normalem]{ulem}
\usepackage{framed}
\usepackage{float}

\makeatletter

\@addtoreset{equation}{section}
\makeatother

\usepackage{parskip}

\def\sqr#1#2{{\vcenter{\vbox{\hrule height.#2pt
\hbox{\vrule width.#2pt height#1pt \kern#1pt \vrule width.#2pt}\hrule
height.#2pt}}}}

\def\b{\beta}

\def\d{\delta}\def\D{\Delta}
\def\e{\epsilon}

\def\k{\kappa}
\def\l{\lambda}
\def\m{\mu}
\def\n{\nu}
\def\r{\rho}
\def\s{\sigma}
\def\t{\tau}

\def\be{\begin{equation}}\def\ee{\end{equation}}
\def\bea{\begin{eqnarray}}\def\eea{\end{eqnarray}}
\def\nn{\nonumber}

\newcommand{\eq}[1]{(\ref{#1})}

\def\tr{{\rm tr}}
\def\Tr{{\rm Tr}}
\newcommand{\w}[1]{\\[0.#1cm]}

\newcommand{\cL}{\mathcal{L}}
\newcommand{\G}{\Gamma}
\def\tv{\Tilde{v}}

\begin{document}

\thispagestyle{empty}

\begin{flushright}\small
MI-HET-867  \\
USTC-ICTS/PCFT-25-39

\end{flushright}

\bigskip

\begin{center}

{ \Large \bf 4D de Sitter from 6D gauged supergravity \\
with Green-Schwarz counterterm}

\end{center}

\vskip 4mm

\begin{center}

{\large Xu Guo$^{\dagger,}$\footnote{xuguo@tamu.edu}, Yi Pang$^{\spadesuit ,\clubsuit }$\footnote{\tt pangyi1@tju.edu.cn}, and Ergin Sezgin$^{\dagger,}$\footnote{\tt sezgin@tamu.edu}}\\
\vskip 8mm

\vskip 3mm

$^\dagger$\,{\it George P. and Cynthia W. Mitchell Institute \\for Fundamental
Physics and Astronomy \\
Texas A\&M University, College Station, TX 77843-4242, USA}\w4

$^\spadesuit $\, {\it Center for Joint Quantum Studies and Department of Physics,\\
 School of Science, Tianjin University, Tianjin 300350, China}
 \w4
$^\clubsuit $\,{\it Peng Huanwu Center for Fundamental Theory, Hefei, Anhui 230026, China}

\end{center}

\vskip1.5cm

\begin{center} {\bf Abstract } \end{center}

Taking into account the Green-Schwarz anomaly counterterm in R-symmetry gauged $N=(1,0)$ supergravity in six dimensions, and the associated modification in the Maxwell kinetic term and potential, the theory admits half-supersymmetric Mink$_4\times S^2$ and non-supersymmetric dS$_4 \times S^2$ solutions with or without a monopole on $S^2$. The monopole charge and the anomaly coefficients play key roles in the vacuum structure and a diagonal gauging in which an admixture of an external $U(1)$ with the R-symmetry $U(1)_R$ is needed for the de Sitter solutions to exist. We determine the full Kaluza-Klein spectrum for both vacua. The spectrum is unitary in the Minkowski case but in the case of de Sitter vacuum, the dilaton and the breathing mode are tachyonic. We show that turning on small perturbations of the tachyonic modes around dS$_4\times S^2$ with a monopole on $S^2$ triggers a flow evolving towards Mink$_4\times S^2$ with the minimal potential energy. The diagonally gauged model also supports dS$_2\times S^4$ solution which avoids tachyons for certain values of the flux on dS$_2$. We also find nonsupersymmetric (A)dS$_4\times S^2$ solutions, when we turn on a flux associated with external $U(1)$ gauge field only, and show that they support the phenomenon of scale separation under certain conditions on anomaly coefficients. 

\setcounter{footnote}{0}

\newpage

\tableofcontents

\vspace{1cm}

\section{Introduction}

It is well known that there are obstacles in obtaining de Sitter solution to string theory under fairly general assumptions \cite{Gibbons:1984kp,Maldacena:2000mw,Dasgupta:2014pma,deWit:1986mwo,Dine:1985he,Andriot:2019wrs,Andriot:2026lac}. To get around this problem, fluxes, branes, orientifolds, $\alpha'$ corrections and quantum effects have been considered in several papers. See\cite{Danielsson:2018ztv,Schachner:2025vol,Cicoli:2023opf} for reviews, and references therein. These attempts have been primarily in a top-down approach, which aims at string/M theory origins. An alternative approach may be a bottom-top approach in which one considers a higher dimensional theory with no known string/theory origin as yet, but satisfying various consistency conditions and at the same time evading the de Sitter no-go theorems. If the search for de Sitter vacua succeeds in a higher dimensional theory, one can then study in depth whether it may be embedded in string/M theory by a novel mechanism which may have been overlooked so far. 

With this motivation in mind, in this paper we consider the vacuum solutions of R-symmetry gauged $N=(1,0)$ supergravity coupled to matter in such a way that it is anomaly free by Green-Schwarz mechanism\footnote{Consistency criteria arise also from anomaly inflow considerations in $6D, (1,0)$ theories but this has been studied only in the absence of R-symmetry gauging so far \cite{Kim:2019vuc,Hamada:2025vga,Brady:2025zzi}.}
A key point in this study is the inclusion of the Green-Schwarz anomaly counterterm associated with the $U(1)_R$ gauge anomaly, which also requires by supersymmetry the modification of the gauge field kinetic terms and the potential, in studying the vacua of the theory. To our best knowledge, this is the first such investigation. In this paper, we shall focus on compactifying solutions to 4D, and we shall not include the gravitational anomaly counterterm, whose supersymmetrization is still not available in presence of $U(1)_R$ gauging. This would require the modification of the 3-form field strength by Lorentz Chern-Simons term, introduction of the Riemann-squared term, as well as 4-derivative terms involving the gauge field strength, and further modification in the potential. We shall comment further in Section 7 on these corrections and on why they are suppressed in the description of our solutions provided by the 2-derivative action.

As far as the consequences of the GS counterterm and the associated terms required by supersymmetry for the solutions of the theory are concerned, we are aware of the following two papers, both in the context of $(1,0), 6D$ supergravity, but without $R$-symmetry gauging.  In \cite{Duff:1996cf},  dyonic string solution was found, and in \cite{Grimm:2013fua}, with certain gauged shift symmetries, a supersymmetric solution was found that takes the form of the product of the 4D Minkowski space and a compact internal space \footnote{A version of this model was considered in \cite{Burgess:2024jkx} where the Green-Schwarz counterterm and the associated modification of the Maxwell kinetic term and the potential are neglected. This model numerically was shown to support a warped product of 4D de Sitter space with a compact internal space, and with two co-dimension 2 singularities.}

With the Green-Schwarz counterterm and its supersymmetric extension at the 2-derivative level taken into account, we find that the theory admits dS$_6$, (A)dS$_4\times S^2$, (A)dS$_2\times S^4$ with no supersymmetry, and Mink$_4\times S^2$ with half-supersymmetry as vacuum solutions. In some of these solutions 2-form flux is needed on either $S^2$ or dS$_2$, and in the first case the monopole quantization condition plays a key role in the result. Remarkably, a diagonal gauging in which an admixture of an external $U(1)$ with the R-symmetry $U(1)_R$, as well as the terms required by the supersymmetrization of the Green-Schwarz counterterm are needed for many of these solutions. Furthermore, in most of the solutions the coefficient of the pure gauge anomaly in the 8-form anomaly polynomial must have particular sign, and their value must be large for the higher derivative corrections to the Lagrangian to be suppressed. Key aspects of the vacua considered in this paper are summarized in Table 4 in the concluding section, and several models in which the desired gaugings, signs and magnitudes of the gauge anomaly coefficients are provided in Appendix A.

We determine the full Kaluza-Klein spectrum for Mink$_4\times S^2$ and dS$_4\times S^2$. The spectrum is unitary in the Minkowski case but in the case of de Sitter vacuum, two scalars violate the Higuchi unitarity bound. These scalars are tachyonic, and one of them comes from the dilaton, and the other from the internal space volume modulus.
In the case of (A)dS$_4\times S^2$ solution in which the 2-form flux is turned on only in the external $U(1)$ direction, we find that the phenomenon of scale separation is realized provided that certain conditions are satisfied by the anomaly coefficients. For a review of scale separation in string theory, see, for example, \cite{Coudarchet:2023mfs}. 

In this paper we have also discuss the solutions (A)dS$_2\times S^4$ and dS$_6$ solutions. In the latter case there are two tachyonic scalars \footnote{Note that the Festina-Lente (FL) bound conjecture proposed in \cite{Montero:2020rpl, Montero:2021otb} assumes the de Sitter vacuum to be stable or metastable, and since the de Sitter vacua found here are all classically unstable, the FL bound is not applicable.} In the case of dS$_2\times S^4$ solution, we have studied the spectrum of the scalar fields and found that the only candidate tachyonic one is absent for a particular value of the $U(1)$ flux on dS$_2$. This unitarity is due to the fact that the linearized equation of motion have a rigid shift symmetry, absent in the Lagrangian with interactions.

As is well known, the presence of tachyons in the context of de Sitter space involving vacua does not mean inconsistency of the theory. Indeed, we show that turning on small perturbations of the tachyonic modes around the dS$_4\times S^2$ in which $U(1)_{R+}$ flux is turned on triggers a flow evolving towards Mink$_4\times S^2$ with the minimal potential energy in the solution space. Interestingly enough, similar tachyons seem to arise in dS$_4$ involving vacuum solutions of Type II string theory effective actions. We refer to \cite{Andriot:2021rdy} where this matter has been discussed in detail with a large number of references. In the context of cosmology, tachyonic scalar fields have been used to drive the inflation (see \cite{Feinstein:2002aj} for instance).
%

\section{The model, anomalies and vacuum solutions}

 In this section, we shall go through the anomaly cancellations in detail because the signs of the anomaly coefficients are crucial for the solutions we will present. There are many papers which use different conventions, and that can lead to a confusion about the interpretation of the signs in the final results.

\subsection{Field content and anomalies}

The $N=(1,0)$ super-Poincar\'e algebra in $6D$ admits the following multiplets
\be
\underbrace{ \left(e_\m^m, \psi_{\m A}^+, B_{\m\n}^{+} \right)}_{\rm graviton}\ ,\qquad \underbrace{\left(B_{\m\n}^{-}, \chi_{A}^-,\varphi  \right)}_{\rm tensor}\ ,\qquad \underbrace{\left(A_\m, \lambda_{A}^+\right)}_{\rm vector}\ ,\qquad \underbrace{\left(4\phi,\psi^- \right)}_{\rm hyper}\ .
\label{fc}
\ee
The two-form potentials, $B_{\m\n}^{\pm}$, have (anti-)self-dual field strengths. The spinors are symplectic Majorana-Weyl, the R-symmetry doublet index $A=1,2$, and fermion chiralities are denoted by $\pm$. The spacetime signature is $(-+++++)$ and the chirality gamma matrix is 
\be
\Gamma_7=\gamma_0\gamma_1\cdots \gamma_5\ .
\ee
Instead of symplectic Majorana-Weyl spinors $\psi^A_\pm$ , we can equivalently work with Weyl spinors $\psi_\pm$. The rest of our conventions are given in Appendix A. We shall consider the couplings of supergravity to $n_T=1$ tensor multiplet, $n_V$ vector multiplets and $n_H$ hypermultiplets. The hyperscalars must parametrize a quaternionic K\"ahler manifold by supersymmetry. We shall assume that this manifold has an isometry group $G_0$. 

Using one of the vector multiplets to gauge the $U(1)_R$ subgroup of the R-symmetry group $Sp(1)_R$, generically there will be gravitational, gauge and mixed anomalies, encoded in an $8$-form anomaly polynomial $I_8$.  Consider a gauge group of the form 
$G_1\times \cdots \times G_n\times U(1)_R$, where  $\prod_{i=1}^n G_i$ is semi-simple. The gravitini, tensorini and gaugini all have charge one under $U(1)_R$. Some or all of the group factors $G_i$ may be subgroups of the isometry group $G_0$. For anomaly freedom, it is necessary that the fatal $\tr\,R^4$ terms in $I_8$ cancel, which imposes the condition
\be 
n_H=n_V+244\ .
\ee
As explained in detail in \cite{Bilal:2008qx}, the variation of the one-loop effective action in Euclidean spacetime denote by $\Gamma_E$ is the anomaly ${\cal A}$ given by
\be
\delta \Gamma_E =-i\int I_6^1 \equiv {\cal A}\ ,
\ee
and the variation of the one-loop effective action in Minkowski spacetime denoted by $\Gamma_M$ is given by 
\be
\delta \Gamma_M = \int I_6^1\ ,
\label{md}
\ee
where $I_6^1$ is obtained from the anomaly polynomial $I_8$ by descent equations
\be
I_8 = dI_7^0\ ,\qquad  \delta I_7^0=dI_6^1\ .
\ee
In the conventions of \cite{Alvarez-Gaume:1984zlq} which we use (with the exception that we take the field strengths to be hermitian), it is important to note that in Euclidean space, the chirality gamma matrix $\Gamma_7^E$ is defined as
\be
\Gamma_7^E = -i\Gamma_1\Gamma_2\cdots \Gamma_6\ .
\ee
Given the rule $i\Gamma^0_M=\Gamma^1_E, \Gamma_M^1=\Gamma_E^2,...,\Gamma_M^5=\Gamma_E^6$, it follows that 
\be
\Gamma_7^E=-\Gamma_7^M \equiv -\Gamma_7\ ,
\label{ir}
\ee
where the label $M$ is short for Minkowski.  Furthermore, in Euclidean space, 
\be
I_8= 2\pi \times \mbox{ index density}\ ,
\ee
where for {\it Euclidean space positive chiral} complex spinors and Euclidean self-dual real 2-form the index is given by \cite{Alvarez-Gaume:1984zlq}
\begin{align}
{\rm ind}\, iD_{3/2} &= \frac{1}{(16\pi^2)^2} \int \left( \frac{245}{360} \tr R^4 -\frac{43}{288} (\tr R^2)^2\right) {\rm dim}\, r -\frac{19}{6} \tr R^2\,\tr_r F^2 +\frac{10}{3} \tr_r F^4\ ,
\nn\w2
{\rm ind}\, iD_{1/2} & = \frac{1}{(16\pi^2)^2} \int  \left( \frac{1}{360} \tr R^4 +\frac{1}{288} (\tr R^2)^2\right) {\rm dim}\, r + \frac16 \tr R^2\,\tr_r F^2  +\frac{2}{3} \tr_r F^4\ ,
\nn\w2
{\rm ind}\, iD_A &= \frac{1}{(16\pi^2)^2} \int \left( \frac{28}{360} \tr R^4 -\frac{8}{288} (\tr R^2)^2\right)\ ,
\end{align}
where ${\rm dim}\, r$ is the dimension of the representation $r$ carried by the spinor. The integral is over Euclidean space in which $ix^0=z^1$. In view of \eq{ir}, the above expressions correspond to {\it Minkowski space negative chiral} complex spinors. Thus, considering the coupling of $n_T=1$ tensor, $n_V$ vector and $n_H$ hypermultiplets, with Minkowski space chiralities and dualities as indicated in \eq{fc}, we obtain the total anomaly polynomial 
\begin{align}
 \frac{1}{2\pi}I_8 &= \frac{1}{(16\pi^2)^2}\Bigg\{(tr\,R^2)^2 -\frac16 \tr\,R^2 \Big[(n_V-20) F^2 + \sum_{i=1}^n \left( \Tr_{ad}\,F_i^2 -\tr_r F_i^2 \right) \Big]
\nn\w2
&+\frac23 \Big[-(n_V+4) F^4 + \sum_{i=1}^n \left( -\Tr_{ad}\,F^4_i + \tr_r\,F_i^4 -6 F^2  \Tr_{ad} \, F^2_i \right)
\nn\w2
& +6\sum_{i<j, r,s} n_{rs}^{ij} ( \tr_r F_i^2)( \tr_s F_j^2) \Big] \Bigg\}\ ,
\label{P}
\end{align}
where $F$  and $F_i$ are the field strengths for $U(1)_R$ and $G_i$ while $\Tr_{ad}$ is the trace in the adjoint representation, $\tr_r$ is the trace in the representation $r$ of the hyperini and $n_{rs}^{ij}$ is the number of hyperfermions carrying the representation $r$and $s$ of the group factors $G_i$ and $G_j$ simultaneously. In our conventions $F_i$ and $F$ are {\it hermitian}.
In an anomaly free model the $\tr\, F^4$ terms either factorize or cancel, and the anomaly polynomial factorizes. In that case it is convenient to write it as
\be
\frac{1}{2\pi} I_8 = \frac12 \eta_{\alpha\beta} Y^\alpha Y^\beta\ ,\qquad \eta_{\alpha\beta} =\begin{pmatrix} 0 & 1\\ 1 & 0 \end{pmatrix}\ ,
\label{P2}
\ee
where the 4-form $Y^\alpha $ is given by
\be
Y^\alpha = \frac{1}{16\pi^2} \left( \frac12 a^\alpha \tr\,R^2 +\sum_{z=1}^{n+1} \frac{2 b_z^\alpha}{\lambda_z} \tr_z\,F^2\right)\ ,
\ee
where $z=1,...,n+1$ labels the factors in the group 
\be
G= \prod_{i=1}^n G_i\times U(1)_R\ ,
\ee
and the traces are in the fundamental representations, and $a^\alpha, b_z^\alpha$ are the constant anomaly coefficients such that
\be
b_z^\alpha =  \begin{cases}
b_i^\alpha & \mbox{for}\quad i=1,...,n \\
b_{n+1}^\alpha = b^\alpha_R & \mbox{for}\qquad U(1)_R 
\end{cases}
\label{bc}
\ee
and $\lambda_z$ are fixed such that the smallest topological charge of an embedded $SU(2)$ instanton is $1$, and they are listed below \cite{Bergshoeff:1985ya, Feger:2019tvk},
\medskip
\begin{align}
&& A_n && B_n && C_n && D_n && E_6 && E_7 && E_8 && F_4 && G_2 
\nn\\
\lambda && 1 && 2 && 1 && 2 && 6 && 12 && 60 && 6 && 2
\label{cn}
\end{align}
with $\lambda_z =1$ in for abelian gauge groups. The anomaly obtained from \eq{md} in Minkowski signature is canceled by the gauge variation of the Green-Schwarz term given by\footnote{In converting the conventions of \cite{Bossard:2024ffp} to ours, we let $B\to \sqrt2 B, A \to 2A$ and $(v,\tv) \to (v,\tv)/(2\sqrt2)$. 
}
\be
\cL_{GS} = \frac18 \e^{\mu\nu\rho\sigma\lambda\tau} B_{\mu\nu} \,\tv^z\,\Tr_z \left(F_{\rho\sigma} F_{\lambda\tau}\right)\ ,
\ee
and the gauge transformation by 
\be
\delta B_{\mu\nu} = 2 v^z \Tr_z \left(\Lambda \partial_{[\mu} A_{\nu]}\right)\ ,
\ee
where $Tr_z (XY)= X^I Y^J \delta_{IJ}$ and $\tv^z= (\tv^i, \tv)$ are constant to be related to the anomaly coefficients.   The gauge variation of $\cL_{GS}$ gives 
\be
\delta \cL_{GS}=-  v_z \tv_z' \Tr_z (\Lambda dA) \,\Tr_{z'} F^2\ ,
\ee
where we have used the relation $dx^{\mu_1} \wedge \cdots\wedge dx^{\mu_6}  = -\epsilon^{\mu_1\cdots \mu_6}d^6 x$. Anomaly freedom requires that
\be
I_6^1 + \delta \cL_{GS}=0\ ,
\ee
which is satisfied for 
\be
\boxed{v_z = \frac{1}{\sqrt{32\pi^3}} b_z^+}\ ,\qquad \boxed{\tv_z = \frac{1}{\sqrt{32\pi^3}} b_z^-}\ ,
\label{vb}
\ee
where $b^\alpha= (b^+, b^-)$. This result provides the relation between the anomaly coefficients and the parameters occurring in the supergravity model.

\subsection{Diagonal gauging}

We see from \eq{P} and \eq{P2} that the class of models described above always have 
\be
U(1)_R\ \mbox{gauged models}: \qquad v\tv<0\ .
\ee
On the other hand, as will be shown in the next section, while the Minkowski$_4\times S^2$ vacuum solution exists for either sign of $v\tv$, the existence of (A)dS$_4\times S^2$ vacuum solutions require that
\be
(A)dS_4 \times S^2:\qquad v\tv >0\ .
\ee
To achieve this, consider the quaternionic K\"ahler symmetric space coset $Sp(n_H,1)/(Sp(n_H)\times Sp(1)_R$, and gauge the diagonal combination of  a $U(1)\in Sp(n_H)$ and $U(1)_R \in Sp(1)_R$, as was done in \cite{Suzuki:2005vu}. This is referred to as \mbox{\it ``diagonal gauging"}. Following \cite{Suzuki:2005vu}, let $\{h_i\}$ be the basis of Cartan subalgebra $u(1)^{n_H}$ of $Sp(n_H)$, then the generator of a $u(1) \in 
u(1)^{n_H}$ is written as
\be
T=\sum_{i=1}^{n_H} q^i h_i\ ,
\label{T}
\ee
where $q^i$ are integers. We shall consider two such models below. 

\subsubsection{ $G=U(1)_{R+}$ model} 

In this case there are $n_H=1+244=245$ hyperini.  We will choose the $U(1)$ generator as 
\be 
T=p(h_1+h_2+....+h_{245})\ .
\label{tmod1}
\ee
Thus we have assigned the $U(1)_{R+}$ charge $p$ to each one of the $n_H=245$ hyperini. Furthermore the $U(1)_{R+}$ generator is the diagonal sum $T^3+T$, where $T_3$ is the $U(1)_R \in Sp(1)_R$ generator.

One finds from \eq{P} that the anomaly polynomial $I_8$ for this model this given by
\be
I_8=\frac{2\pi}{(16\pi^2)^2}\left[({\rm tr}R^2)^2+\frac16(19+245p^2){\rm tr}R^2 F^2+\frac23(245p^4-5)F^4\right]\ .
\ee
In order to have $v\tv>0$, we need $p\ge1$.  In particular, when $p=1$, we have 
\begin{align}
\frac{1}{2\pi }I_8 &=\frac1{(16\pi^2)^2}\left({\rm tr}R^2+4 F^2\right)\left({\rm tr}R^2+40 F^2\right) =\frac12 \eta_{\alpha\beta} Y^\alpha Y^\beta\ ,
\label{m1}
\end{align}
where 
\be
Y^\alpha = \frac{1}{16\pi^2} \left( \frac12 a^\alpha \tr R^2 +2 b_{R+}^{\alpha} F^2\right)\ .
\label{dgY}
\ee
In the convention of \cite{Monnier:2017oqd}, we read off the properly normalized SO(1,1) anomaly vector associated with the Lorentz group and $U(1)_{R+}$
\be
a^\alpha=(2,\, 2),\quad b_{R+}^\alpha=(2,20)\ , 
\ee
from which we can read of $v^\alpha$ by using \eqref{vb}. Clearly, the vectors $a$ and $\frac12 b_{R+}$ lie on an even unimodular lattice with basis vectors
\be
e_1^{\alpha}=(1,\,0),\quad e_2^{\alpha}=(0,\,1)\ .
\ee

\subsubsection{$G=U(1)_{R+}\times U(1)^\prime$ model}

In this case we take the gauge group to have an additional $U(1)$. Thus we have $n_H=2+244=246$. Considering a particular model considered in \cite{Suzuki:2005vu}, the number of hyperini carrying charges $(\pm p,\pm q)$ under $U(1)_{R+}\times U(1)^\prime$ is denoted by $n_{pq}$ given by
\be
n_{pq} = \begin{pmatrix} 2&4&6\\150&4&2\\6&62&10\ 
\end{pmatrix}\ .
\ee
Thus the $U(1)$ generator $T$ and $U(1)'$ generator $T'$ are given by
\begin{align}
T &= \sum_{n=1}^{12} (-1)^n h_n +\sum_{n=13}^{168} 2(-1)^n h_n +\sum_{n=169}^{246} 3(-1)^n h_n \ ,
\label{tm1}\w2
T' &= \sum_{n=1}^{2} (-1)^n h_n +2\sum_{n=3}^{6} (-1)^n h_n +
+ 3\sum_{n=7}^{12} (-1)^n h_n
\nn\w2
& + \sum_{n=13}^{162} (-1)^n h_n + 2\sum_{n=163}^{166} (-1)^n h_n + 3\sum_{n=167}^{168} (-1)^n h_n
\nn\w2
& +\sum_{n=169}^{174} (-1)^n h_n
+2 \sum_{n=175}^{236} (-1)^n h_n
+3 \sum_{n=237}^{246} (-1)^n h_n\ .
\label{tp}
\end{align}
The alternating signs in the sums means that a half of them have charge $(p,q$)and the other half $(-p,-q)$, so that the $F F_+^3$ and $F^3 F_+$ terms in the anomaly polynomial cancel out \cite{Suzuki:2005vu}. With the above charge assignments, the resulting anomaly polynomial factorizes as  
\be
\frac{1}{2\pi} I_8=\frac{1}{(16\pi^2)^2}\left({\rm tr}R^2+30 F^2+76F^{'2}\right)\left({\rm tr}R^2+196 F^2+24F^{'2}\right)=\frac12 \eta_{\alpha\beta} Y^\alpha Y^\beta\ ,
\label{m2}
\ee
where $F$ and $F'$ are the $U(1)_{R+}$ and $U(1)^\prime$ field strengths and 
\be
Y^\alpha=\frac{1}{16\pi^2}\left(\frac12 a^\alpha{\rm tr}R^2 +b_1^\alpha F^{'2}+ b^\alpha_{R+} F^2\right)\ ,
\ee
with
\be
a^\alpha=(2,\, 2),\quad b^\alpha_{R+}=(15,98),\quad b_1^\alpha=(38,12)\ ,
\label{m2ac}
\ee
where $a$, $\frac12 b_{R+}$ and $\frac12 b_1$ must lie on certain unimodular lattice. We find that such a lattice indeed exists with basis vectors 
\be
e_1^{\alpha}=(1,\,0),\quad e_2^{\alpha}=(\frac12,\,1)\ .
\ee
Note also that  using \eq{bc} and \eq{vb}, we can deduce $v^\alpha$ and $v^{\prime\alpha}$ from \eq{m2ac}. This is another example of a model with $v\tv>0$.

\subsection{The bosonic Lagrangian and field equations}

We now consider the bosonic part of the Lagrangian describing the two models with diagonal gauging described above. In presence of the Green-Schwarz anomaly counterterm, a general pseudo-Lagrangian is given in \cite{Riccioni:2001bg}, and for the case of single tensor multiplet for which there is a proper Lagrangian, the result described in \cite{Bossard:2024ffp} will be used. Leaving aside the hypermultiplet couplings that will not play a role for the solutions we shall present, and denoting the gauge group by $G=\prod_z G_z = U(1)_{R+} \times U(1)'$,  the bosonic part of the proper Lagrangian is given by \cite{Bossard:2024ffp} 
\begin{align}
e^{-1}\mathcal{L} = &\frac{1}{\k^2} \Big[ \frac14 R - \frac14\partial_\mu \phi \partial^\mu \phi -\frac{1}{12} e^{2\phi} H_{\mu \nu \rho} H^{\mu \nu \rho} -\frac14\k f_z F_{z\mu\nu}F_z^{\mu\nu} 
\nonumber \\
& +\frac18 e^{-1} \k \tv_z\,
\epsilon^{\mu \nu \rho \sigma \tau \lambda} B_{\mu \nu} F_{z\rho \sigma} F_{z\tau\lambda} 
+\frac12 e^{-1}\epsilon^{\mu\nu\rho\sigma\lambda\tau} \k^2 v_z \tv_{z'} X_{z\mu\nu\rho}X_{z'\sigma\lambda\tau} -V \Big]\ ,
\label{Lagrangian}
\end{align}
where $H_{\mu\nu\rho} = 3\partial_{[\mu} B_{\nu\rho]} - 6 \k v_z\,X_{z\mu\nu\rho}$ with  $X_{\mu\nu\rho}:=A_{[\mu}\partial_\nu A_{\rho]}$ and $X_{1\mu\nu\rho}:=A^\prime_{[\mu}\partial_\nu A_{\rho]}^\prime$, and 
\be
f_z=(f, f_1)\ ,\qquad v_z=(v_{R+}, v_1)\ ,\qquad f= v_{R+} e^{\phi} + \tv_{R+} e^{-\phi}\ ,\qquad
f_1=v_1 e^\phi + \tv_1 e^{-\phi}\ ,
\ee
with $v_z^\alpha$ and $\tv_z^\alpha$ from \eq{vb} and \eq{m2ac}. {\it In the rest of the paper we set $(v,\tv)= \left(v_{R+}, \tv_{R+}\right)$, for simplicity in notation.}
The potential for the two models under consideration is given by 
\begin{align}
U(1)_{R+}\ \mbox{model:}\qquad & V=-\frac{1}{\kappa} \frac{\tr\ C^2}{f}\ ,
\label{mod1}\w2
U(1)_{R+} \times U(1)'\ \mbox{model:} \qquad & V= -\frac{1}{\kappa} \frac{\tr\,C^2}{f} - \frac{1}{\kappa} \frac{\tr\ C_1^2}{f_1}\ ,
\label{mod2}
\end{align}
and the $C$-functions are defined as \cite{Randjbar-Daemi:2004bjl}
\begin{align}
C_{AB} &= \Big[ L^{-1}\left(T^3+T\right)L\Big]_{AB}\ ,
\label{CC}\w2
C_{1AB} &= \left(L^{-1} T^\prime L\right)_{AB}\ ,
\label{C1}
\end{align}
where $L$ is a representative of the coset $Sp(n_H,1)/(Sp(n_H)\times Sp(1)_R)$, $T^3=-\frac12 i \sigma^2$ is the $U(1)_R\in Sp(1)_R$ generator, $T$ is defined in \eq{tmod1}, and $T'$ is the $U(1)\in Sp(n_H)$ generator. The key point for the vacua we will be describing is that we can set the hyperscalars $\varphi=0$ consistent with their equation of motion, and doing so gives \footnote{Note that $\tr C^2= \tr (T^3)^2 = -1/2$ when we set the hyperscalars equal to zero, which means a positive definite potential for $f>0$, in agreement with \cite{Salam:1984cj}.}
\be 
\tr\,C^2|_{\varphi=0} =\frac12 \ ,\qquad C_1\Big|_{\varphi=0} =0\ .
\ee

The mass dimensions are $[\phi]=[B_{\m\n}]=0, [A_\m]=1, [\kappa]=-2, [v]=[\tv]=0$. Below, except in conclusions, we shall set $\k=1$ for simplicity in notation. Rescaling the metric as $g_{\mu\nu}= e^\phi g_{\mu\nu}'$ shows that $\langle e^{-2\phi}\rangle$ plays a role similar to the string loop expansion parameter in heterotic string. In the terms other than the potential, the terms not involving $v$ are tree level, and those involving $\tv$ are 1-loop terms. The potential term in string frame is of the schematic form
\be
V\propto \frac{\sqrt{-g} e^{2\phi}}{v+\tv e^{-2\phi}}\ ,
\label{pot}
\ee
which in the weak coupling limit, contains infinite loop contributions characterized by power of $e^{-2\phi}$. Even though the potential in the Einstein frame, $\sqrt{-g}/f$, is consistent with $S$-duality exchanging $v\to \tv$ and $\phi\to-\phi$, whether it admits a nonperturbative interpretation is not known \footnote{A potential of the form  $\sqrt{-g} C^2(\varphi)/f$ where $C(\varphi)$ is a function of the hyperscalars (different in nature than ours) obtained from gauging of a shift symmetry in the context of F-theory compactification on Calabi-Yau manifold to 6D, but apparently its (non)perturbative interpretation is not known in that case either. }.

In the case of a single $U(1)_{R+}$ symmetry gauging, the Lagrangian above enjoys the following scaling symmetry
\be
g_{\m\n}\to g_{\m\n},\quad B_{\m\n}\to \lambda B_{\m\n},\quad A_{\m}\to A_{\m},\quad e^{\phi}\to \lambda^{-1} e^{\phi},\quad v\to \lambda v,\quad \tv\to \lambda^{-1} \tv\ .
\ee
We see that the field equations below are invariant under this symmetry transformation, and thus the spectrum depends on the scaling invariant product $v\tv$. 

The bosonic field equations derived from the Lagrangian \eq{Lagrangian}, with hyperscalars consistently set to zero, are given by
\begin{align}
\Box\phi\: =&  \: \frac13 e^{2\phi} H^2 + \frac12 f'_z Tr_zF^2 +(f^{-1})'\ ,
\label{Dilaton}\w2
R_{\mu\nu} =&  \: \partial_\mu \phi \partial_\nu \phi +
e^{2\phi} ( H^2_{\mu\nu} - \frac{1}{6}H^2 g_{\mu\nu} ) +
2 f_z (Tr_zF^2_{\mu\nu}-\frac{1}{8} g_{\mu\nu} Tr_zF^2 ) +\frac12 f^{-1} g_{\mu\nu}\ ,
\label{Einstein}\w2
D_\mu\left( f_z F_z^{\mu\nu } \right) =&
-v_z e^{2\phi} H^\nu_{\ \rho\sigma}F_z^{\rho\sigma} -
\frac16\tv_z\, \epsilon^{\nu\mu abcd} H_{\mu ab} F_{z cd}
\nn\\
&-\frac12 (v_z\tv_{z'}+\tv_zv_{z'})\epsilon^{\nu abcde} X_{z'abc}  F_{z de} 
-\frac18 (v_z\tv_{z'}+\tv_zv_{z'})\epsilon^{\nu abcde} A_{z a} Tr_z'(F_{bc}F_{de})\ ,
\label{Maxwell} \w2
\nabla_{[\mu} H_{\nu\rho\sigma]} =&-\frac32 v_z Tr_z (F_{[\mu\nu} F_{\rho\sigma]})\ ,
\label{Bianchi}\w2
\nabla_{\mu}(e^{2\phi}  H^{\mu\nu\rho}) =& -\frac14 \tv_z \epsilon^{\nu\rho\sigma\tau\lambda\theta}Tr_z (F_{\sigma\tau}F_{\lambda\theta})\ , 
\label{H eom}
\end{align}
where $f'_z := \frac{\partial f_z}{\partial\phi}$ and we have set $\kappa=1$ which can be reintroduced by dimensional analysis. In \eq{Maxwell}, there is summation over $z'$ but not $z$. Since the hyperscalars do not play a role for the kind of vacua we are interested in, they are consistently set to zero. 
The displayed Green-Schwarz term is associated with the gauge anomalies. Gravitational and gauge-gravitational mixed anomalies would introduce Lorentz Chern-Simons modification of the 3-form field strength and 4-derivative extension of the model, including Riemann-squared terms as well as $F^4$ terms and modifications in the potential. For the vacuum solutions we shall present and study here, these terms will be suppressed under certain conditions, in particular the requirement that $v\tv\gg1$, which in turn translate to large internal space volume, as discussed in Section 7. 

Recalling that we are setting the hypermultiplet sector to zero, the supertransformations take the form
\begin{align}
\delta \psi_\mu &= D_\mu \epsilon + 
\frac{1}{24} e^\phi H_{\nu\rho\sigma}\gamma^{\nu\rho\sigma}\gamma_\mu \epsilon 
\ ,
\\
\delta \chi &= \frac12 \gamma^\mu \partial_\mu \phi \epsilon - \frac{1}{12} e^\phi \gamma^{\mu\nu\rho}H_{\mu\nu\rho} \epsilon \ ,
\\
\delta \lambda &= - \frac{1}{4} \gamma^{\mu\nu}F_{\mu\nu}\epsilon +\frac{i}{2} f^{-1} \epsilon \ ,
\\
\delta \lambda^\prime &= - \frac{1}{4} \gamma^{\mu\nu}F^\prime_{\mu\nu}\epsilon \ ,
\end{align}
where $f=ve^\phi+\tv e^{-\phi}$ and
\be
D_\mu \epsilon = \partial_\mu \epsilon +\frac{1}{4} \omega_\mu^{\ ab}\gamma_{ab}\epsilon - i A_\mu \epsilon\ ,
\ee
where $\e=\e_+$ is a Weyl spinor. In obtaining this result, the fact that the shift term in $\delta \lambda \sim f^{-1} C \Big|_{\varphi=0} \sim f^{-1}$ and in $\delta\lambda' \sim (f_1)^{-1}C_1\Big|_{\varphi=0}=0$ has been used.

\subsection{The Mink\texorpdfstring{$_4 \times S^2$}{4 x S2} and dS\texorpdfstring{$_4 \times S^2$}{4 x S2}  solutions}
\label{sec:Mink4}

We take the 6D spacetime to be a direct product of two constant curvature spaces $M_4 \times M_2$, and turn on only $U(1)_{R+}$ flux on $S^2$. More precisely, splitting the indices as $\mu,\nu =0,1, 2,3$, $i,j = 4,5$, we make the ansatz
\begin{align}
& R_{ \mu \nu \rho \sigma } = \frac{\epsilon}{L^2} ( g_{ \mu \rho } g_{ \nu \sigma } -g_{ \mu \sigma } g_{ \nu \rho } )\ ,\quad 
R_{ i j k l }= \frac{1}{a^2} ( g_{ i k } g_{ j l }-g_{ i l }g_{ j k } ), 
\nn\w2
& F_{ij} = k\, \epsilon_{ij}\ ,\quad \phi = \phi_0\ ,\quad 
 f_0 \equiv v e^{\phi_0} + \tv e^{-\phi_0} >0\ ,\quad \mbox{rest}=0\ ,
\label{bg}
\end{align}
where $\e,L, a, k$ are constants. In order that the Maxwell field is globally well defined on $S^2$, the associated 2-form flux must be quantized as
\begin{align}
ka^2 = \frac{1}{2}n \qquad,\qquad n\in\mathbb{Z}\ .
\label{qc}
\end{align}
We find the following solutions, valid for both $U(1)_{R+}$ and $U(1)_{R+}\times U(1)'$ models:
\be
\boxed{\mbox{Mink}_4\times S^2}\qquad 
\e=0\ ,\quad \phi = \phi_0\ ,\quad k = \pm \frac{1}{f_0} \ ,\quad  \frac{1}{a^2} = \frac{2}{f_0}>0\ , 
\label{susy solution}
\ee
This solution is half-supersymmetric and the quantization condition \eq{qc} is satisfied since $ka^2= \pm 1/2$. 
Putting aside the quantization we also have dS$_4\times S^2$ solutions for which 
\begin{align}
& \boxed{\mbox{dS}_4 \times S^2}  \quad \e=+1\ ,\quad
\phi_0  = \frac12\ln\left(\frac{\tv}{v}\right)\ ,
\quad k a^2 = \pm \frac{1}{2}\sqrt{1+\frac{6}{\beta^2}-\frac{27}{\beta^4}}\ ,
\label{dsc}
\end{align}
where
\begin{align}
\beta^2 \equiv\frac{L^2}{a^2}=\frac{36k^2 v\Tilde{v}+3}{-4k^2 v\Tilde{v}+1}\ .
\end{align}
Using the quantization condition \eq{qc} in \eq{dsc} gives
\be
\frac{1}{\beta^2} = \frac19 \left(1\pm \sqrt{4-3n^2}\right)\ ,\qquad n \in \mathbb{Z}\ .
\ee
Thus, we must have $n=\pm1$ or $n=0$, corresponding to the following dS$_4\times S^2$ solutions
\begin{align}
n=0:& \quad \beta^2=3\ , \quad L^2=12 \sqrt{v\tv} =3 a^2\ ,\quad k=0\ ,\quad \phi_0=\frac12 \ln \left(\frac{\tv}{v}\right)\ ,
\label{S1}\w2
n=\pm 1: &\quad  \beta^2=\frac92\ ,\quad L^2=\frac{27}{2} \sqrt{v\tv}=\frac92 a^2\ ,\quad k=\pm \frac{1}{6\sqrt{v\tv}} \ ,\quad \phi_0=\frac12 \ln \left(\frac{\tv}{v}\right)\ .
\label{S2}
\end{align}
These de Sitter solutions are nonsupersymmetric. Furthermore, for both of them the relation  $f'|_{\phi=\phi_0}=0$ holds.

\section{Linearized field equations and harmonic expansions on \texorpdfstring{$S^2$}{S2}} 

In this section we shall perform harmonic expansions on $S^2$. In this process we shall pick gauge fixing conditions accommodated by gauge symmetries and supersymmetry. Even though there are gauge and supersymmetry anomalies at the classical level, they vanish for the vacuum solutions we are considering. To be more specific, the gauge symmetry and supersymmetry holds for the quadratic action that results for perturbations around these vacua. The detailed structure of the anomalies are discussed in detail in \cite{Ferrara:1997gh}, and the statements made above can easily be checked by making use of the anomaly expressions provided there.

\subsection{The bosonic sector}

We parameterize the fluctuations around the background \eq{susy solution}, \eqref{S1} and \eqref{S2} is follows
\begin{align}
&g_{MN} = \bar{g}_{MN} + h_{MN}\ , \quad \phi = \phi_0 + \varphi,\quad A_M = \bar{A}_M + a_M\ ,\quad B_{MN} = \bar{B}_{MN} + \Tilde{b}_{MN}\ .
\end{align}
We make field redefinition $b_{MN} \equiv \Tilde{b}_{MN}+2v\bar{A}_{[M}a_{N]}$, such that we have the linearized fluctuation
\begin{align}
\delta H_{MNP} = 3\partial_{[M}b_{NP]}-6v \overline{F}_{[MN}a_{P]}\ ,
\end{align}
and under the $U(1)_R$ gauge transformation, the 2-form $b_{MN}$ transform as 
\be
\delta a_M=\partial_M\Lambda,\quad \delta b_{MN}=2v\bar{F}_{MN} \Lambda\ .
\ee
About the background solution, other linearized gauge symmetries are \cite{Pang:2012xs}
\bea
\delta h_{MN}&=&\bar{\nabla}_M\xi_N+\bar{\nabla}_N\xi_M,\quad \delta a_M=\xi^N\bar{F}_{NM}\ ,
\nn\\
\delta b_{MN}&=&\partial_M\Lambda_N-\partial_N\Lambda_M\ . 
\eea
On $S^2$ the transverse spin-1 harmonics $Y^{(\ell)}_i$ are related to the spin-0 harmonics $Y^{(\ell)}$ by
\begin{align}
Y^{(\ell)}_i = \epsilon_i^{\ j}\nabla_j Y^{(\ell)}\ ,
\end{align}
and they satisfy 
\begin{align}
\Box_2 Y^{(\ell)} = -\frac{\ell(\ell+1)}{a^2} Y^{(\ell)}\quad,\quad \Box_2 Y^{(\ell)}_i = -\frac{\ell(\ell+1)-1}{a^2} Y^{(\ell)}_i\ .
\end{align}
We choose the de Donder-Lorentz gauge:
\begin{align}
\nabla^i h_{ij} =\frac{1}{2}\nabla_j h^k_{\ k} \ ,\quad \nabla^i h_{i \mu}=0\ ,\quad\nabla^i a_i =0\ ,\quad \nabla^i b_{iM} = 0 .
\end{align}
Harmonic expansions under this gauge take the form
\begin{align}
\varphi & = \sum_{\ell\geq 0} \phi^{(\ell)} Y^{(\ell)}\ ,
h_{\mu\nu}  = \sum_{\ell\geq 0} h^{(\ell)}_{\mu\nu} Y^{(\ell)}\ ,
h_{\mu i} = \sum_{\ell\geq 1}  h^{(\ell)}_\mu Y^{(\ell)}_i \ ,
h_{ij}  = g_{ij}\sum_{\ell\geq 0} N^{(\ell)} Y^{(\ell)}\ ,
\nonumber\\
a_\mu & = \sum_{\ell\geq 0} a^{(\ell)}_\mu Y^{(\ell)}
\ , \  
a_i = \sum_{\ell \geq1} a^{(\ell)}Y^{(\ell)}_i 
\ , \  
b_{\mu\nu} = \sum_{\ell\geq 0} b^{(\ell)}_{\mu\nu} Y^{(\ell)} \ ,\ 
b_{\mu i} = \sum_{\ell\geq 1} b^{(\ell)}_\mu Y^{(\ell)}_i \ ,
\nonumber\\
b_{ij} & =\epsilon_{ij} b^{(0)}Y^{(0)} \ ,
\label{exps}
\end{align}
where the expansion coefficients, e.g. $\Phi^{(\ell)}$ are functions of the 4D spacetime. The Donder-Lorentz gauge does not fix all the gauge symmetries, and consequently there are some residual ones generated by harmonic zero modes, namely the $S^2$ Killing vector $Y^{(1)}_i$ and conformal Killing vectors $\nabla_i Y^{(1)}$. These residual gauge symmetries are:
\begin{itemize}
\item The 4D coordinate transformation generated by $ \xi_\mu =\xi^{(0)}_\mu Y^{(0)}$:
\begin{align}
\delta h^{(0)}_{\mu\nu} = \nabla_\mu \xi^{(0)}_\nu + \nabla_\nu \xi^{(0)}_\mu\ .\label{ctg}
\end{align}
\item The Stuckelberg shift symmetries generated by $\xi_\m = -\nabla_\m\xi^{(1)} Y^{(1)},\,\xi_i = \xi^{(1)}\nabla_i Y^{(1)}$:
\be
\delta h_{\m\n}^{(1)}=-2\nabla_\m\nabla_\n\xi^{(1)},\quad
\delta N^{(1)}=-\frac{2}{a^2}\xi^{(1)},\quad \delta a^{(1)}=-k\xi^{(1)}\ .
\label{sss}
\ee
\item Linearized $SU(2)$ symmetry generated by $\xi_i=\xi'^{(1)}Y^{(1)}_i$, $\Lambda=-k\xi'^{(1)}Y^{(1)}$ and
$\Lambda_i=-vk\xi'^{(1)}Y^{(1)}_i$\ .
\be
\delta h_{\m}^{(1)}=\nabla_{\m} \xi'^{(1)},\quad\delta a^{(1)}_{\m}=-k\nabla_\m\xi'^{(1)},\quad \delta b^{(1)}_{\m}=vk^2a^2\nabla_\m\xi'^{(1)}\ .
\ee
\item Four dimensional $U(1)_R$ symmetry generated by $\Lambda = \Lambda^{(0)}Y^{(0)}$:
\begin{align}
\delta a^{(0)}_\mu = \partial_\mu \Lambda^{(0)},\quad \delta b^{(0)}=2vk\Lambda^{(0)}\ .\label{urs}
\end{align}
\item Abelian 2-form symmetry generated by $\Lambda'_\mu = \Lambda''^{(0)}_\mu Y^{(0)}$:
\be
\delta b^{(0)}_{\m\n}=\partial_\m \Lambda''^{(0)}_\n-\partial_\n \Lambda''^{(0)}_\m\ .
\ee
\end{itemize}
Performing the harmonic expansions on the 2-sphere, we find the linearized field equations given below where we write $f$ for $f_0$ for simplicity in notation. It is also understood that in the case of $\epsilon=0$ we have in mind the solution \eq{susy solution}. Next we list the linearized field equations in which the harmonic expansions on $S^2$ have been carried out.

\subsection*{\it Dilaton field equation }

%
\begin{align}
(\Box_4 - M^2_\ell) \phi^{(\ell)} 
+ \frac{6\epsilon}{L^2} \phi^{(\ell)} 
-\frac{1-\mid\epsilon\mid}{a^2}\left(\frac{f'^2}{f^2}\phi^{(\ell)}
+\frac{f'}{f} M^2_\ell a^{(\ell)}
-\frac{f'}{f} N^{(\ell)}\right)=0\ ,\quad (\ell \geq 0)\ ,
\label{d31}
\end{align}
where
\be
M^2_\ell := \frac{\ell(\ell+1)}{a^2}\ .\label{ml}
\ee

\subsection*{\it Einstein equations}
%
\begin{align}
&\left( \Box_4 - M^2_\ell - \frac{2\epsilon}{L^2} \right)h^{(\ell)}_{\mu\nu} 
-2\nabla_{(\mu} \nabla^\rho h^{(\ell)}_{\nu)\rho} 
+2\nabla_\mu \nabla_\nu N^{(\ell)} \nonumber 
+ 2g_{\mu\nu} k^2 f N^{(\ell)} 
\\
&\quad - 2 g_{\mu\nu} k f  M^2_\ell a^{(\ell)} 
-g_{\mu\nu}\frac{1-\mid\epsilon\mid}{a^2}\frac{f'}{f}\phi^{(\ell)}= 0\ ,\quad 
(\ell \geq 0)\ ,
\label{en31}
\\
&\left(\Box_4 - M^2_\ell + \frac{3\epsilon}{L^2} \right) h^{(\ell)}_\mu 
- 4 k f a^{(\ell)}_\mu =0\ ,
\quad (\ell \geq 1)\ ,
\label{en32}
\\
&
\nabla^\nu h^{(\ell)}_{\mu\nu} - \nabla_\mu N^{(\ell)} + 4 k f \nabla_\mu a^{(\ell)} = 0 \ ,
\quad (\ell \geq 1)\ ,
\label{en33}
\\
&\left(\Box_4 - M^2_\ell + \frac{6\epsilon}{L^2} - 2 k^2 f \right)N^{(\ell)}+ 6k f M^2_\ell a^{(\ell)} 
+ \frac{1-\mid \epsilon \mid}{a^2}\frac{f'}{f}\phi^{(\ell)}= 0\ ,
\quad (\ell \geq0)\ ,
\label{en34}
\\
&\nabla^\mu h^{(\ell)}_\mu = 0\ , \quad (\ell \geq 1)\ ,
\label{en35}
\\
& h^{(\ell)\mu}_\mu = 0\ ,
\quad (\ell \geq1) \ .
\label{en36}
\end{align}

\subsection*{\it Maxwell equations}
%
\begin{align}
& \left( \Box_4 - M^2_\ell - \frac{3\epsilon}{L^2} 
- \frac{4v^2 k^2 e^{2\phi}}{f} \right) a^{(\ell)}_\mu 
- k M^2_\ell h^{(\ell)}_\mu 
+ \frac{k\Tilde{v}}{f}\epsilon_{\mu\nu\rho\sigma}\nabla^\nu b^{(\ell)\rho\sigma}
\nonumber \\
&\quad - 2 v e^{2\phi_0} k\frac{1}{f} M^2_\ell b^{(\ell)}_\mu 
+ 2\delta^\ell_0 v e^{2\phi_0} k \frac{1}{f}\nabla_\mu b^{(0)}= 0\ , \quad (\ell \geq0)\ ,
\label{a31}
\\
&(\Box_4 - M^2_\ell ) a^{(\ell)} + k N^{(\ell)} 
-(1-\mid \epsilon \mid)k\frac{f'}{f} \phi^{(\ell)}= 0\ ,\quad
(\ell \geq 1)\ ,
\label{a32}
\\
&\nabla^\mu a^{(\ell)}_\mu = 0 \ , \quad (\ell \geq 1)\ .
\label{a33}
\end{align}
\subsection*{\it Tensor field equations}
%
\begin{align}
&\left( \Box_4 - M^2_\ell -\frac{4\epsilon}{L^2}\right)b^{(\ell)}_{\mu\nu} + 2 k\Tilde{v}e^{-2\phi} \epsilon_{\mu\nu\rho\sigma}\nabla^\rho a^{(\ell)\sigma} = 0\ ,\quad (\ell \geq0)\ ,
\label{b31}\\
&\left( \Box_4 - M^2_\ell - \frac{3\epsilon}{L^2} \right)b^{(\ell)}_\mu - 2 vka^{(\ell)}_\mu = 0\ ,
\quad (\ell \geq 1)\ ,
\label{b32}\\
&\nabla^\nu b^{(\ell)}_{\mu\nu} = 0 \ , \quad (\ell \geq 0)\ ,
\label{b33}
\w2
&
\nabla^\mu b^{(\ell)}_\mu = 0 \ ,\quad (\ell\geq1)\ ,
\label{b35}
\w2
&\Box_4 b^{(0)} - 2vk \nabla^\mu a^{(0)}_\mu= 0\ .
\label{b34}
\end{align}

\subsection{The fermionic sector}

For the Minkowski$_4\times S^2$ and (A)dS$_4\times S^2$ vacua, the linearized field equations for the fermions simplify considerably and take the form
\begin{align}
& \Gamma^{ABC} D_B \psi_C +\frac12 f \Gamma^{BC}\gamma^A \lambda F_{BC}- i \Gamma^A \lambda =0 \ ,
\label{gravitino1}\w2
& \Gamma^A D_A\chi -\frac12 f^\prime F_{AB}\Gamma^{AB}\lambda -i\frac{f^\prime}{f}\lambda =0\ ,
\label{dilatino1}\w2
&f \Gamma^A D_A \lambda +\frac14 f F_{BC}\Gamma^A \Gamma^{BC}\psi_A +\frac14 f^\prime F_{AB}\Gamma^{AB} \chi -\frac{i}{2}\Gamma^A\psi_A +\frac{i}{2}\frac{f^\prime}{f} \chi=0\ ,
\label{gaugino1}
\end{align}
and the supertransformations for the solutions in which $H=0$ and $\phi={\rm constant}$, are
\begin{align}
\delta \psi_A &= \left( \partial_A +\frac{1}{4} \omega_A^{\ MN}\Gamma_{MN} - i A_A \right) \epsilon\ ,\label{susyf1}
\\
\delta \lambda &= - \frac{1}{4} \Gamma^{AB}F_{AB}\epsilon +\frac{i}{2} f^{-1} \epsilon \ ,\label{susyf2}
\\
\delta \chi &=0\ .\label{susyf3}
\end{align}
In the equations above, the spinors are Weyl, as opposed to symplectic Majorana-Weyl.
We choose
\begin{align}
\Gamma^a &=\gamma^a\times \sigma^3 \ ,\qquad \Gamma^i=1\times \sigma^i\  ,\quad (i=1,2)\ .
\end{align}
Thus, we have $\Gamma_7=\Gamma_0\Gamma_1...\Gamma_6 = \gamma_5 \times \sigma^3$ and 
\begin{align}
    \Gamma_7 \psi_M = \psi_M \quad,\quad
    \Gamma_7 \chi = - \chi \quad,\quad
    \Gamma_7 \lambda = \lambda\ .
\end{align}
%


Here we shall first assume that $k>0$, and therefore $s=ka^2+\frac12=1$. We shall comment on the case of $k<0$ at the end. In this sector, we need the gauge covariant derivative in the presence of a Dirac monopole with potential $A=-ka^2\cos\theta d\phi$,  one can use the scalar spin-weighted harmonics ${}_{s}Y_{\ell m}$ to build the spinor harmonics as \cite{Pang:2012xs}
\be
k>0, s=1: \qquad \eta^{(\ell)}_+= \begin{bmatrix}
{}_{0}Y_{\ell m} \\
 0  
\end{bmatrix}\ , \quad \mbox{for}\ \ell\ge 0\ , \quad \eta^{(\ell)}_-= \begin{bmatrix}
0 \\
i({}_{1}Y_{\ell m} )
\end{bmatrix}
\ , \quad \mbox{for}\ \ell \ge 1\ , \quad \ell=0,1,2,...
\label{fep}
\ee
which obey 
\be
\s^i D_i\eta^{(\ell)}_\pm=i\frac{1}{a}\sqrt{\ell(\ell+1)}\eta^{(\ell)}_\mp =iM_{\ell}\, \eta^{(\ell)}_\mp \ ,
\ee
on $S^2$.
When $\ell=0$, we only have $\eta^{(0)}_+$ since ${}_{1}{Y_{0,m}}$ does not exist.
Next, we choose the gauge 
\begin{align}
\psi_{\{i\}} = 0 \label{fgauge}\ ,
\end{align}
where $\{i\}$ means $\Gamma$-traceless. The harmonic expansion under this gauge gives
\begin{align}
&\psi_\mu  =  \psi^{(0)}_{\mu+} \otimes \eta^{(0)}_+ + \sum_{\ell>0} \left(\psi^{(\ell)}_{\mu+} \otimes \eta^{(\ell)}_+ + \psi^{(\ell)}_{\mu-} \otimes \eta^{(\ell)}_-\right)\ ,
\\
&\psi_i  = \Gamma_i \Psi^{(0)}_- \otimes \eta^{(0)}_+ + \sum_{\ell>0} \Gamma_i \left(\Psi^{(\ell)}_{-} \otimes \eta^{(\ell)}_{+} + \Psi^{(\ell)}_+ \otimes \eta^{(\ell)}_-\right)\ ,
\\
&\chi = \chi^{(0)}_- \otimes \eta^{(0)}_+ + \sum_{\ell>0} \left(\chi^{(\ell)}_+ \otimes \eta^{(\ell)}_- +  \chi^{(\ell)}_- \otimes \eta^{(\ell)}_+ \right)\ ,
\\
&\lambda = \lambda^{(0)}_+ \otimes \eta^{(0)}_+ + \sum_{\ell>0} \left( \lambda^{(\ell)}_+ \otimes \eta^{(\ell)}_+ + \lambda^{(\ell)}_- \otimes \eta^{(\ell)}_-\right)\ .
\end{align}
Note that $\psi_{\mu +}$ and $\psi_{\mu -}$ are independent Weyl spinors in 4D. The gauge choice \eq{fgauge} does not fix all the gauge symmetries, and we find the following residual symmetry transformations
\begin{itemize}
\item Generated by ${\epsilon}^{(0)}=\epsilon^{(0)}_+\otimes\eta^{(0)}_+$:
\begin{align}
\delta\psi^{(0)}_{\mu+} = \nabla_\mu \epsilon^{(0)}_+\ ,
\ \delta \lambda^{(0)}_+=-\frac{i}{2}\left(k-\frac{1}{f}\right)\epsilon^{(0)}_+\ .
\label{rsf1}
\end{align}
\item Generated by $\epsilon^{(1)}=\epsilon^{(1)}_-\otimes\eta^{(1)}_-$:
\begin{align}
\delta \psi^{(1)}_{\mu-}=\nabla_\mu \epsilon^{(1)}_-\ ,\ 
\delta \Psi^{(1)}_-=i\frac{\sqrt{2}}{2a} \epsilon^{(1)}_-\ ,\ 
\delta \lambda^{(1)}_-=\frac{i}{2}\left(k+\frac{1}{f}\right)\epsilon^{(1)}_-\ .
\label{rsf2}
\end{align}
\end{itemize}
We shall take these symmetries into account in the spectrum analysis. Next, we list the all the linearized fermionic field equations.

\subsection*{\it Gravitino equations\ $(\ell>0)$  }
%
\begin{align}
&\gamma^{\mu\nu\rho}\nabla_\nu \psi^{(\ell)}_{\rho+} 
- i M_\ell \gamma^{\mu\nu} \psi^{(\ell)}_{\nu-} +2\gamma^{\mu\nu}\nabla_\nu \Psi^{(\ell)}_- 
+iM_\ell \gamma^\mu \Psi^{(\ell)}_+ 
+i (kf-1) \gamma^\mu \lambda^{(\ell)}_+
= 0\ ,
\label{3/2+1}\\
&\gamma^{\mu\nu\rho}\nabla_\nu \psi^{(\ell)}_{\rho-} 
+iM_\ell\gamma^{\mu\nu} \psi^{(\ell)}_{\nu+} 
-2\gamma^{\mu\nu}\nabla_\nu \Psi^{(\ell)}_+ 
+iM_\ell\gamma^\mu\Psi^{(\ell)}_- 
-i(kf+1)\gamma^\mu\lambda^{(\ell)}_- 
= 0\ ,
\label{3/2+2}\\
&\gamma^{\mu\nu}\nabla_\mu \psi^{(\ell)}_{\nu+}
-i\frac{1}{2}M_\ell\gamma^\mu\psi^{(\ell)}_{\mu-}
+\gamma^\mu \nabla_\mu \Psi^{(\ell)}_- 
-i (kf+1)\lambda^{(\ell)}_+ = 0\ , \quad\label{3/2+3}
\\
&\gamma^{\mu\nu}\nabla_\mu \psi^{(\ell)}_{\nu-}
+i\frac{1}{2}M_\ell\gamma^\mu\psi^{(\ell)}_{\mu+}
-\gamma^\mu \nabla_\mu \Psi^{(\ell)}_+ 
+i(kf-1)\lambda^{(\ell)}_-
= 0
\ ,\quad\label{3/2+4}
\end{align}
\subsection*{\it Dilatino equations \ $(\ell>0)$  }
%
\begin{align}
& \gamma^\mu \nabla_\mu \chi^{(\ell)}_+ 
- iM_\ell\chi^{(\ell)}_- 
-i\frac{f'}{f}(kf-1)\lambda^{(\ell)}_-
= 0\ ,
\label{dilatino11}
\\
&\gamma^\mu \nabla_\mu \chi^{(\ell)}_- 
+ iM_\ell\chi^{(\ell)}_+     
-i\frac{f'}{f} (kf+1)\lambda^{(\ell)}_+
=0\ .
\label{dilatino2}
\end{align}
\subsection*{\it Gaugino equations \ $(\ell>0)$  }
%
\begin{align}
& \gamma^\mu \nabla_\mu \lambda^{(\ell)}_+ 
+ i M_\ell\lambda^{(\ell)}_- 
+ i\frac{1}{2f}(kf-1)\gamma^\mu \psi_{\mu+}
- i\frac{1}{f}(kf+1)\Psi^{(\ell)}_- 
+ i \frac{f'}{2f^2}(kf+1)\chi^{(\ell)}_-
=0\ ,
\label{gaugino11}
\\
& \gamma^\mu \nabla_\mu \lambda^{(\ell)}_- 
- i M_\ell\lambda^{(\ell)}_+ 
- i\frac{1}{2f}(kf+1)\gamma^\mu \psi_{\mu-}
- i\frac{1}{f}(kf-1)\Psi^{(\ell)}_+ 
+ i \frac{f'}{2f^2}(kf-1)\chi^{(\ell)}_+
=0\ .
\label{gaugino2}
\end{align}

When $\ell=0$, four physical modes vanish, since they need the spin-weighted harmonic function ${}_sY_{\ell m}$ with $\ell=0$, $s=1$, which does not exist.
The zero modes fermions are $(\psi^{(0)}_+\ ,\ \Psi^{(0)}_-\ ,\ \chi^{(0)}_-\ ,\ \lambda^{(0)}_+)$, and their field equations are 
\begin{align}   
&\gamma^{\mu\nu\rho}\nabla_\nu\psi^{(0)}_{\rho+}+2\gamma^{\mu\nu}\nabla_\nu \Psi^{(0)}_- + i (kf-1)\gamma^\mu \lambda^{(0)}_+ = 0\ ,
\label{fz1}\\
&\gamma^{\mu\nu}\nabla_\mu \psi^{(0)}_{\nu+}+\gamma^\mu\nabla_\mu\Psi^{(0)}_- -i(kf+1)\lambda^{(0)}_+=0 \ ,
\label{fz2}\\
&\gamma^\mu\nabla_\mu \chi^{(0)}_--i\frac{f'}{f} (kf+1)\lambda^{(0)}_+=0\ ,
\label{fz3}\\
&\gamma^\mu \nabla_\mu\lambda^{(0)}_+ -i\frac{1}{f}(kf+1)\Psi^{(0)}_- + i\frac{1}{2f}(kf-1)\gamma^\mu \psi^{(0)}_{\mu+}+ i \frac{f'}{2f^2}(kf-1)\chi^{(0)}_+=0\ .
\label{fz4}
\end{align}
So far, we assumed that $k>0$ which implied $s=1$. If we take $k<0$, it gives $s=0$ in the spin weighted harmonics ${}_sY_{\ell m}$ and the only difference in the equations given for the case of $k>0$ is the flipping of the chiralities for the $\ell=0$ modes.

\subsection*{The case of $k=0:$}

For the case with $k=0$, $s=\frac12$, the harmonic expansion of $\eta_\pm^{(\ell)}$ has the same form as in \eqref{fep} but with 
\begin{align}
\ell=\frac12,\frac32,\dots\ ,
\end{align}
Accordingly, one may use the expansion
\be
\eta^{(\ell)}_+= \begin{bmatrix}
{}_{-\frac12}Y_{\ell m} \\
0  
\end{bmatrix},\quad \eta^{(\ell)}_-= \begin{bmatrix}
0 \\
i{}_{\frac12}Y_{\ell m}  
\end{bmatrix}
\ ,\  \label{fep0}
\ee  
and
\begin{align}
\s^iD_i\eta^{(\ell)}_\pm=\s^i\nabla_i \eta^{(\ell)}_\pm=i\frac1a\left(\ell+\frac12\right) \eta^{(\ell)}_\mp \ .
\end{align}
In this case, the linearized field equations take the same form as \eqref{3/2+1}-\eqref{gaugino2}. Unlike the previous case, however, there are no $\ell=0$ modes. Moreover, the gauge choice \eq{fgauge} does not fix all the gauge symmetries, one finds the following residual symmetry transformations generated by ${\epsilon}^{(1/2)}=\epsilon^{(1/2)}_+\otimes\eta^{(1/2)}_++\epsilon^{(1/2)}_-\otimes\eta^{(1/2)}_-$:
\begin{align}
\delta\psi^{(1/2)}_{\mu\pm} = \nabla_\mu \epsilon^{(1/2)}_\pm\ ,
\ \delta \lambda^{(1/2)}_\pm=\frac{i}{2f}\epsilon^{(1/2)}_\pm\ .
\label{rsf3}
\end{align}
We shall take these residual symmetries into account in the spectrum analysis. Regrading all the other equations given in the case of $k\ne 0$, the only other difference is to replace $M^2_\ell= \ell(\ell+1)/a^2$ for $\ell \in \mathbb{Z}$ with $M^2_\ell= \left(\ell+\frac12\right)^2/a^2$ for $\ell\in \mathbb{Z}+\frac12$. 

\section{Spectrum on \texorpdfstring{$\mbox{Mink}_4\times S^2$}{Mink4 x S2}}

In this section, we will use the formulas presented in Sections 2 and Section 3 by setting $\epsilon=0$ and keeping $\phi_0$ arbitrary. Next, we go to the momentum space in Minkowski$_4$ and determine the masses. In our spacetime signature, the tachyon-free free condition is simply that the momentum-squared is non-negative on-shell. We begin by treating the $\ell \ge 1$ modes and then consider the $\ell=0$ modes separately. 
The resulting spectrum is for supersymmetric Minkowski$_4\times S^2$ solution, and we will see that for the special value $\tv=0$ it agrees with the result given long ago in \cite{Salam:1984cj}.

\subsection{Bosonic sector}

At the level of the equations of motion, we can readily check whether there exists a tachyon in the spectrum. 

\subsubsection*{\boxed{\it Spin\, 2}}

From \eq{en31}, by using the spin-2 projector we get
\begin{align}
(\Box_4 -M^2_\ell)h^{TT(\ell)}_{\mu\nu} =0\ ,\qquad \ell \ge 0\ ,
\end{align}
where $h^{TT}$ denotes the transverse-traceless projection. Thus the masses of the spin 2 tower are given by 
\be
m^2_\ell = \ell(\ell+1)/a^2\ , 
\label{spin2 mass}
\ee
which are all non-tachyonic. The $\ell=0$ mode is the massless graviton.

\subsubsection*{\boxed{\it Spin\, 1}}

{\it The $\ell\ge 1$ modes:} 

This sector consists of $(\partial^\mu h_{\mu\nu}^T\ ,\ \partial^\mu b_{\mu\nu}\ ,\ h^T_\mu\ ,\ a^T_\mu\ ,\ b^T_\mu\ ,\ b^T_{\mu\nu})$. Acting with $\partial^\mu$ on \eq{en32} and \eq{b31}, we get
\begin{align}
(\partial^\mu h^{(\ell)}_{\mu\nu})^T =0\ ,\quad \partial^\mu b^{(\ell)}_{\mu\nu} =0\ ,\qquad \ell \ge 1\ .
\end{align}
For the other four spin 1 fields, we need to solve \eq{en32}, \eq{b31}, \eq{a31} and \eq{b32}. To handle the curl of fields in the equations, we act with $\epsilon_{\alpha\beta\gamma\mu}\partial^\mu$ on \eq{en32}, \eq{a31} and \eq{b32}, obtaining for $\ell\ge1$,
\begin{align}
&(\Box_4-M^2_\ell)*F_{\mu\nu}(h^{(\ell)})-4*F_{\mu\nu}(a^{(\ell)})=0\ ,
\w2
&(\Box_4-M^2_\ell)b^{(\ell)T}_{\mu\nu} 
+\frac{\Tilde{v}}{a^2}e^{-2\phi_0}*F_{\mu\nu}(a^{(\ell)})=0\ ,
\w2
&(\Box_4 - M^2_\ell)*F_{\mu\nu}(b^{(\ell)})-\frac{v}{a^2}*F_{\mu\nu}(a^{(\ell)}) = 0\ ,
\w2
&\left(\Box_4-M^2_\ell-\frac{v^2 e^{2\phi_0}}{2a^6}\right)*F_{\mu\nu}(a^{(\ell)}) 
-\frac{M^2_\ell}{2a^2}*F_{\mu\nu}(h^{(\ell)})  
-M^2_\ell \frac{ve^{2\phi_0}}{2a^4} *F_{\mu\nu}(b^{(\ell)})  
+\frac{\Tilde{v}}{2a^4}\Box_4 b^{(\ell)}_{\mu\nu} = 0\ .
\end{align}
Here, $*F(a)$ is the dual of the field strength of $a$. In momentum space, the condition for these equations to have a nontrivial solution gives the following eigenvalues for the masses for $\ell\ge 1$: 
\begin{align}
m^2_{(i),\ell} &= \frac{\ell(\ell+1)}{a^2}\ , \quad  i=1,2
\label{spin1 mass0}
\\
m^2_{\pm,\ell} &= \frac{\ell(\ell+1)+1}{a^2} - \frac{v\tv}{2a^6} 
\pm \frac{1}{a^2}\left[4\ell(\ell+1)\left(1-\frac{v\tv}{4a^4}  \right) + \left(1-\frac{v\tv}{2a^4}\right)^2\right]^\frac12\ .
\label{spin1 mass}
\end{align}
It follows that for $\ell >1$, the combinations of $(h^T_\mu\ ,\ a^T_\mu\ ,\ b^T_\mu\ ,\ b^T_{\mu\nu})$  describe 4 towers of massive spin 1 fields with masses given above. Note that massive $b_{\mu\nu}$ is on-shell dual to a  massive vector in 4D. 
The masses for $m^2_{\pm,\ell}$ are real for $v\tv>0$ provided that $a^2 \ge \sqrt{v\tv}$. With this condition satisfied, a straightforward analysis also shows that there are no tachyons. 

For $\ell=1$, we observe that $m^2_{-}=0$, which means that there is a combination of $(h^T_\mu,\ a^T_\mu,\ b^T_\mu,\ b^T_{\mu\nu})$ that describes a triplet of massless vector fields. 

{\it The $\ell=0$ modes:} 

This sector contains $b^{(0)T}_{\mu\nu}$ and $a^{(0)T}_\mu$. By choosing $U(1)_R$ gauge $\partial^\mu a^{(0)}_\mu = \frac{a^2}{v}\Box_4 b^{(0)}$ supported by the symmetry \eq{urs}, and defining $c_\mu \equiv a^{(0)}_\mu +\frac{v}{a^2}\partial_\mu b^{(0)}$, we get from \eq{b31} and the action of $\epsilon_{\alpha\beta\gamma\mu}\partial^\mu$ on \eq{a31} the following equations 
\begin{align}
&\Box_4 b^T_{\mu\nu}  + \frac{\Tilde{v}}{a^2}e^{-2\phi_0}*F_{\mu\nu}(c) = 0\ ,
\label{argue1}
\\
&\left(\Box_4-\frac{v^2 e^{2\phi_0}}{2a^6}\right) *F_{\mu\nu}(c) 
+ \frac{\Tilde{v}}{2a^4}\Box_4 b^T_{\mu\nu} = 0 \ ,
\label{argue2}
\end{align}
where $F(c)=dc$. The masses of the two fields are obtained as
\begin{align}
m^2 = 0\quad ,\qquad m^2=\frac{2}{a^2} \left(1-\frac{v\tv}{2a^4}\right)\ .
\end{align}
These happen to agree with the extrapolation of the formula for $m^2_-$ given \eq{spin1 mass} to $\ell=0$. In more detail, the massless mode arises in this sector as follows: Defining $\Tilde{b}_{\mu\nu} := b^T_{\mu\nu} + M^2_0\tv e^{-2\phi_0}/a^2 *F_{\mu\nu}(c)$, where $M^2_0 := \frac{2}{a^2}\left(1-\frac{v\tv}{2a^4}\right)$, \eq{argue1} and \eq{argue2}take the form
\begin{align}
\Box_4 \Tilde{b}_{\mu\nu}=0\quad,\quad 
(\Box_4- M^2_0)*F_{\mu\nu}(c)=0\quad,
\end{align}
The first equation describes a massless 2-form field which is dual to massless scalar field in $4D$. The second equation is what results from the action of $\epsilon^{\mu\nu\rho\sigma}\partial_\sigma$ on the Proca equation $(\Box_4-m^2) A_\rho=0$.

\subsubsection*{\boxed{\it Spin\, 0}}

{\it The $\ell\ge 1$ modes:} 

This sector contains $(\partial^\mu h^{(\ell)}_\mu\ ,\ \partial^\mu a^{(\ell)}_\mu\ ,\ \partial^\mu b^{(\ell)}_\mu\ ,\ h^{T(\ell)}\ ,\ h^{L(\ell)}\ , \ N^{(\ell)}\ ,\ a^{(\ell)}\ ,\ \phi^{(\ell)})$, here, $h^{T(\ell)} :=\frac{1}{3}T^{\mu\nu}h^{(\ell)}_{\mu\nu}$ and $h^{L(\ell)}:= L^{\mu\nu}h^{(\ell)}_{\mu\nu}$. From \eq{en35}, \eq{a33} and \eq{b33}, we have for $\ell \ge 1$,
\begin{align}
\partial^\mu h^{(\ell)}_\mu=0\ ,\quad \partial^\mu a^{(\ell)}_\mu=0\ ,\quad \partial^\mu b^{(\ell)}_\mu=0\ .
\end{align}
From \eq{en33}-\eq{en36}, we have for $\ell \ge 1$,
\begin{align}
3h^{T(\ell)} +h^{L(\ell)} = 0\ ,\quad h^{L(\ell)} = N^{(\ell)}-4a^{(\ell)}\ .
\end{align}
Hence, $h^{T(\ell)}$ and $h^{L(\ell)}$ are not independent fields. Thus, we are left with $(N^{(\ell)},a^{(\ell)},\phi^{(\ell)})$ as independent fields. The remaining 3 scalars obey the field equations for $\ell\ge 1$,
\begin{align}
&\left(k^2 +M^2_\ell+\frac{1}{a^2}\right)N^{(\ell)}_\ell - 6M^2_\ell a^{(\ell)} - \frac{1}{a^2}\frac{f'}{f}\phi^{(\ell)} =0\ ,
\\
&-\frac{1}{2a^2} N^{(\ell)} +(k^2+M^2_\ell)a^{(\ell)} + \frac{1}{2a^2} \frac{f'}{f}\phi^{(\ell)} = 0\ ,
\\
&-\frac{1}{a^2}\frac{f'}{f}N^{(\ell)} + 2M^2_\ell \frac{f'}{f} a^{(\ell)} + \left(k^2 + M^2_\ell + \frac{1}{a^2}\frac{f'^2}{f^2} \right)\phi^{(\ell)}=0\ .
\end{align}
The determinant roots of this $3\times 3$ matrix gives the masses for $\ell\ge 1$ as
\begin{align}
m^2_\ell &= \frac{\ell(\ell+1)}{a^2} \ ,
\label{spin0 mass0}
\\
m^2_{\pm,\ell} &= \frac{\ell(\ell+1)+1}{a^2} - \frac{v\Tilde{v}}{2a^6} 
\pm \frac{1}{a^2}\left[4\ell(\ell+1)\left(1-\frac{v\Tilde{v}}{4r^4}  \right) + \left(1-\frac{v\Tilde{v}}{2r^4}\right)^2\right]^\frac12\ .
\label{spin0 mass}
\end{align}
These masses are the same as those of three of the vector fields determined above, consistent with supersymmetry, as we shall see later. 

For $\ell=1$ we observe that $m^2_{-}=0$, which means that there is a combination of $(N^{(1)},a^{(1)},\phi^{(1)})$ that describes a triplet of massless scalar fields. The residual gauge symmetry \eqref{sss} suggests that this massless mode is eaten by the massive spin-2 mode.

{\it The $\ell=0$ modes:} 

Here we have the scalars $(h^{T(0)},\ h^{L(0)},\ N^{(0)},\ \phi^{(0)})$. 
From \eq{en36} and the gauge choice in \eq{ctg}, we have $h^{T(0)} = 0$ and $h^{L(0)}=0$. 
We have two scalars left, and their equations \eq{d31}\eq{en34} are given
\begin{align}
&(\Box_4-\frac{1}{a^2}\frac{f'^2}{f^2})\phi^{(0)}-\frac{1}{a^2}N^{(0)} = 0\ ,
\\
&(\Box_4-\frac{1}{a^2})N^{(0)} + \frac{1}{a^2} \frac{f'}{f}\phi^{(0)} = 0\ ,
\end{align}
The masses of the two fields are
\begin{align}
m^2 = 0 \quad,\qquad m^2 = \frac{2}{a^2}\left(1-\frac{v\tv}{2a^4}\right)\ .
\end{align}
As before, these masses coincide with the extrapolation of \eq{spin0 mass} to $\ell=0$.

\subsection{Fermionic sector}

\subsubsection*{\boxed{\it Spin\, 3/2}}

From \eq{3/2+1} and \eq{3/2+2}, projecting with the spin 3/2 projector we have for $\ell \ge 0$,
\begin{align}
&\gamma^{\mu\nu\rho}\partial_\nu (P^{3/2}\psi)^{(\ell)}_{\rho+} 
- i M_\ell \gamma^{\mu\nu} (P^{3/2}\psi)^{(\ell)}_{\nu-} 
= 0\ ,
\\
&\gamma^{\mu\nu\rho}\partial_\nu (P^{3/2}\psi)^{(\ell)}_{\rho-} 
+iM_\ell\gamma^{\mu\nu} (P^{3/2}\psi)^{(\ell)}_{\nu+} 
= 0\ ,
\end{align}
where $P^{3/2}$ is the spin $3/2$ projector.  Thus, we have two spin 3/2 towers with masses $\ell(\ell+1)/a^2$ for $\ell\ge 1$.

When $\ell=0$, from \eq{fz1}, projecting with the spin 3/2 we have
\begin{align}
\gamma^{\mu\nu\rho}\partial_\nu(P^{3/2}\psi)^{(0)}_{\rho+} = 0\ .
\end{align}
Thus, we have one massless gravitino.

\subsubsection*{\boxed{\it Spin\, 1/2}}

{\it $\ell\ge 1 $ modes:}

There are ten spin 1/2 fields ($\Lambda_\pm, \Omega_\pm, \Psi^{(\ell)}_\pm, \chi^{(\ell)}_\pm, \lambda^{(\ell)}_\pm$ ) for $\ell \ge 1$, where $\Lambda_\pm := \gamma^\mu \psi^{(\ell)}_{\mu\pm}$ and $\Omega_\pm := \partial^\mu \psi^{(\ell)}_{\mu\pm}$. First, we find that ($\Lambda_\pm, \Omega_\pm$) are not physical modes, since the field equations give 
\begin{align}
&\Lambda_\pm = 0\ ,\qquad \Omega_- = -i M_\ell \Psi^{(\ell)}_- +2i \lambda^{(\ell)}_-
\ ,\qquad \Omega_+ = -i M_\ell \Psi^{(\ell)}_+-3i \lambda^{(\ell)}_+ \ .
\end{align}
The remaining six spin-1/2 fields are $( \ \Psi^{(\ell)}_+\ ,\ \Psi^{(\ell)}_-\ ,\ \chi^{(\ell)}_+\ ,\ \chi^{(\ell)}_-\ ,\ \lambda^{(\ell)}_+\ ,\ \lambda^{(\ell)}_-\ )$, and the field equations describing their mixing are for $\ell \ge 1$,
\begin{align}
&\gamma^\mu \partial_\mu \Psi^{(\ell)}_+ - iM_\ell \Psi^{(\ell)}_-  +2i \lambda^{(\ell)}_- = 0 \ , 
\\
&\gamma^\mu \partial_\mu \Psi^{(\ell)}_- + iM_\ell \Psi^{(\ell)}_+ 
 +i \lambda^{(\ell)}_+= 0\ ,
\\
&\gamma^\mu \partial_\mu \chi^{(\ell)}_+ - iM_\ell \chi^{(\ell)}_-
 = 0\ ,
\\
&\gamma^\mu \partial_\mu \chi^{(\ell)}_- 
+ iM_\ell \chi^{(\ell)}_+ - 2i\frac{f'}{f}\lambda^{(\ell)}_+= 0\ ,
\\
&\gamma^\mu \partial_\mu \lambda^{(\ell)}_+ + iM_\ell \lambda^{(\ell)}_-
-i \frac{2}{f}\Psi^{(\ell)}_- 
+ i\frac{f'}{f^2}\chi^{(\ell)}_- = 0
\ ,
\\
&\gamma^\mu \partial_\mu \lambda^{(\ell)}_- - iM_\ell \lambda^{(\ell)}_+
= 0
\ .
\end{align}
By solving the determinant of the $6\times 6$ matrix, we find that the masses are 
\begin{align}
m^2_\ell &= -\frac{\ell(\ell+1)}{a^2}\ ,
\\
m^2_{\pm,\ell} &= \frac{\ell(\ell+1)+1}{a^2} - \frac{v\Tilde{v}}{2a^6} 
\pm \frac{1}{a^2}\left[4\ell(\ell+1)\left(1-\frac{v\Tilde{v}}{4a^4}  \right) + \left(1-\frac{v\Tilde{v}}{2a^4}\right)^2\right]^\frac12\ .
\label{half}
\end{align}

When $\ell=1$, we have triplet of massless spinors since
\begin{align}
m^2_- =0\ ,\quad \mbox{for}\quad \ell=1\ .
\end{align}

{\it $\ell=0$ modes:}

The field equations in this sector are 
\begin{align}
&\gamma^\mu\partial_\mu \Psi^{(0)}_- +i\lambda^{(0)}_+ = 0 \ ,
\\
&\gamma^\mu\partial_\mu\chi^{(0)}_- 
-2i \frac{f'}{f}\lambda^{(0)}_+ = 0\ ,
\\
&\gamma^\mu\partial_\mu \lambda^{(0)}_+ 
-i \frac{2}{f}\Psi^{(0)}_- + i\frac{f'}{f^2}\chi^{(0)}_- =0 \ .
\end{align}
From this we deduce the masses 
\begin{align}
m^2=0\quad,\quad m^2= \frac{2}{a^2}\left(1-\frac{v\Tilde{v}}{2a^4}\right)\ ,
\end{align}
which means a massless Majorana fermion and a single massive Dirac fermion, which counts as two Majorana. 

\subsection{Summary}

The full spectrum for the Mink$_4\times S^2$ vacuum solution is given in Table 1. The dependence on the masses on the parameter $\tv$ arises in the two of the spin-1 towers only. 
In the limit $\tv=0$, the massive spectrum reduces to what was given long ago in \cite{Salam:1984cj}, except that the presence of massless scalar multiplet, and the fact a would-be massless vector multiplet becomes massive was noted later in \cite{Aghababaie:2002be}. Consistent dimensional reduction and the question of whether a chiral $N=1$ theory in 4D can be obtained was studied in \cite{Pope:2011yi}.
\begin{table}
\centering
\small
\begin{tabular}{|c|c|c|}
\hline
 s & $ a^{-2} m^2$ & \mbox{Comments}\\
\hline\hline
$2$  & $\ell(\ell+1)$ & $\ell \ge 0$\\
\hline
$\frac{3}{2}$  & $\ell(\ell+1)$ & $\ell\ge 1$:\quad complex massive gravitino 
\\
&& $\ell=0$: \quad massless gravitino
\\
\hline
& $\ell(\ell+1)$ & $\ell\geq1$
\\
& $\ell(\ell+1)$  &$\ell\geq1$
\\
$1  $&$\ell(\ell+1)+1 - \frac{v\Tilde{v}}{2a^4} 
\pm \left[4\ell(\ell+1)\left(1-\frac{v\Tilde{v}}{4a^4}  \right) + \left(1-\frac{v\Tilde{v}}{2a^4}\right)^2\right]^\frac12$ & $m^2_+$ for $\ell\ge 0$
\\
&&  $m^2_-=0$ for $\ell=0$ (massless 2-form)
\\
&& $m^2_-=0$ for $\ell=1$
\\
\hline
&  $\ell(\ell+1)$ & $\ell \ge 1$:\ complex massive spinor
\\
$\frac12$  &  $\ell(\ell+1)+1 - \frac{v\Tilde{v}}{2a^4} 
\pm \left[4\ell(\ell+1)\left(1-\frac{v\Tilde{v}}{4a^4}  \right) + \left(1-\frac{v\Tilde{v}}{2a^4}\right)^2\right]^\frac12$& $m^2_+$ for $\ell \ge 0$:\ complex massive spinor
\\
&& $m^2_-$ for $\ell \ge 2$:\ complex massive spinor
\\
&& $m^2_-=0$ for $\ell=0,1$
\\
\hline
& $\ell(\ell+1)$ & $\ell \ge 1$ 
\\
$0$ & $\ell(\ell+1)+1 -\frac{v\Tilde{v}}{2a^4} 
\pm \left[4\ell(\ell+1)\left(1-\frac{v\Tilde{v}}{4a^4}  \right) + \left(1-\frac{v\Tilde{v}}{2a^4}\right)^2\right]^\frac12$ & $m^2_+$ for $\ell \ge 0$
\\
&& $m^2_-=0$ for $\ell=0$
\\
\hline
\end{tabular}
\caption{The complete spectrum on Minkowski$_4\times S^2$. The massless modes consist of a supergravity multiplet, a triplet of vector multiplets and one scalar multiplet. The massive KK modes are such that at each KK level $\ell \ge 1$ there is  a massive spin-2 multiplet $(e_\mu^a, 2 \psi_\mu, A_\mu)$ with mass $m_\ell^2=\ell(\ell+1)/a^2$, and two massive vector multiplets each consisting of $(A_\mu, 2\chi, \phi)$ with $m^2_\ell= \ell(\ell+1)/a^2$ and $m_\ell^2= m^2_{+(\ell)}$. At $\ell=0$ there is an additional vector multiplet with mass $m^2_+\vert_{\ell=0}$. At each level $\ell \ge 2$ there is also a massive vector multiplet with $m^2_\ell = m^2_{-(\ell)}$. A triplet of massless spin-1/2 states and a triplet of massless spin-0 states with $m^2_-=0$ arise, which are removed by the residual shift symmetry \eqref{rsf2} and \eqref{sss}, respectively. See \eq{spin1 mass} for $m^2_\pm$.}
\end{table}

\newpage

\section{Spectrum on \texorpdfstring{$\mbox{dS}_4\times S^2$}{dS4 x S2}}

In this section, we will use the formulas presented in Sections 2 and 3 by setting $\epsilon=+1$. What remains to be done in this section is to perform the harmonic expansion on dS$_4$ and determine the conformal dimensions. We shall then check the so-called Higuchi bounds which are needed for unitarity. We begin by treating the $\ell \ge 1$ modes and then consider the $\ell=0$ modes separately. 

In this section, we keep $\beta^2=L^2/a^2$ arbitrary for part of the discussion, since the theory admits two $\mathrm{dS}_4\times S^2$ solutions described in Section~2: the first has non-vanishing monopole flux, $k\neq 0$, for which $\beta^2=9/2$, while the second has vanishing monopole flux, $k=0$, for which $\beta^2=3$. Most of the detailed spectrum analysis below will be presented for the $\beta^2=9/2$ solution. As we shall see in the concluding section, the effective potential arising in $4D$ is everywhere positive for the case of $\beta^2=9/2$ while it is unbounded from below the case of $\beta^2=3$. Nonetheless, we shall determine the full spectrum for both cases.

\subsection{The bosonic sector}

\subsubsection*{\boxed{\it Spin\, 2}}

Only $h^{(\ell)}_{\mu\nu}$ contributes to this sector. Given that $h_{\mu\nu}^{(\ell)}$ is traceless for $\ell\ge 1$, its harmonic expansion in $dS_4$ is given by 
\begin{align}
h^{(\ell)}_{\mu\nu} (x)=\sum_{\Delta,\ell\ge1} h^{(\Delta 2)(\ell)}D^{(\Delta 2)}_{\mu\nu}(x)+\sum_{\Delta,\ell\ge1} h^{(\Delta 1)(\ell)}D^{(\Delta 1)}_{\mu\nu}(x)+\sum_{\Delta,\ell\ge1} h^{(\Delta 0)(\ell)}D^{(\Delta 0)}_{\mu\nu}(x)\ ,
\end{align}
where $D^{(\Delta s)}_{\mu\nu}(x)$ are the $SO(4,1)$ representation functions. Substituting this into \eq{en31}, projecting to the $(\Delta,2)$ representation and using \eq{g3}, we get
\begin{align}
\left(\frac{\Delta(\Delta-3)}{L^2}+\frac{\ell(\ell+1)}{a^2}\right)h^{(\Delta2)(\ell)} = 0\ ,
\label{s2e}
\end{align}
with the solutions 
\begin{align}
\Delta_{2\pm}=\frac32 \pm \frac12 \sqrt{9-4\b^2\ell(\ell+1)}\ , \quad \ell\ge 0\ ,
\label{spin2}
\end{align}
For $\beta^2=9/2$ as well as $\beta^2=3$, the unitary representations involved are the principal series for $\ell\ge 1$ and the exceptional series for $\ell=0$, the latter describing the dS massless graviton.

\subsubsection*{\boxed{\it Spin\, 1}}

{\it The $\ell \ge 1$ modes:} 

When $\ell \geq 1$, in this sector, we have the fields $\left(\nabla^\mu h_{\mu\nu}^{(\ell)}, \nabla^\mu b_{\mu\nu}^{(\ell)}, h^{(\ell)}_\mu, a^{(\ell)}_\mu, b^{(\ell)}_\mu, b^{(\ell)}_{\mu\nu}\right) $. The first field does not contribute to the spin-1 sector since we find from \eq{en33} that $h^{(\Delta1)(\ell)}=0$ for $\ell\ge 1$.  Next, we rescale the remaining fields as
\begin{align}
A^{(\ell)}_\mu:=\frac{1}{k}a^{(\ell)}_\mu
\ , \ 
B^{(\ell)}_\mu:=\sqrt{\frac{\tv}{v}}b^{(\ell)}_\mu
\ , \ 
B^{(\ell)}_{\mu\nu}:=\sqrt{\frac{\tv}{v}}b^{(\ell)}_{\mu\nu}\ .
\end{align}
As can be seen from the field equations, these fields are divergent-free. Therefore, they do not have spin-0 modes and their harmonic expansions on dS$_4$ take the form 
\begin{align}
&h^{(\ell)}_\mu(x)=\sum_\Delta h^{(\Delta 1)(\ell)} D^{(\Delta 1)}_\mu(x)
\quad,\quad
A^{(\ell)}_\mu(x)=\sum_\Delta A^{(\Delta 1)(\ell)} D^{(\Delta 1)}_\mu(x)\ ,
\\
&B^{(\ell)}_\mu(x)=\sum_\Delta B^{(\Delta 1)(\ell)} D^{(\Delta 1)}_\mu(x)
\quad,\quad
B^{(\ell)}_{\mu\nu}(x)=\sum_\Delta C^{(\Delta 1)(\ell)} D^{(\Delta 1)}_{[\mu\nu]+}(x) + \Tilde{C}^{(\Delta 1)(\ell)} D^{(\Delta 1)}_{[\mu\nu]-}(x)\ .
\end{align}
From $\nabla^\mu b^{(\ell)}_{\mu\nu}=0$, it follows that $C^{(\Delta1)(\ell)}=\Tilde{C}^{(\Delta1)(\ell)}$.
Defining
\begin{align}
    x:=\Delta(\Delta-3)+2\ .
\end{align}
The unitary condition can be rewritten as 
\begin{align}
x\leq-\frac{1}{4}& \ ,\qquad \mbox{principle series}\ ,
\\
-\frac{1}{4}<x<0& \ , \qquad \mbox{complimentary series}\ ,
\\
x=0& \ , \qquad \mbox{exceptional series}\ .
\end{align}
The linearized field equations in the spin 1 sector are
\begin{align}
&[x-6+\beta^2 \ell(\ell+1)]h^{(\Delta1)(\ell)}
+2(\beta^2-3)A^{(\Delta 1)(\ell)} = 0\ ,
\\
&\left[x + \beta^2 \ell(\ell+1)+\frac{1}{2}(\beta^2-3)\right]A^{(\Delta 1)(\ell)}+\beta^2 \ell(\ell+1) B^{(\Delta1)(\ell)}
+\beta^2 \ell(\ell+1)h^{(\Delta1)(\ell)}
\nn\\
& +\sqrt{3x}LC^{(\Delta 1)(\ell)}=0\ ,
\\
&[x+\beta^2 \ell(\ell+1)]L C^{(\Delta 1)(\ell)}
-\frac{1}{2}(\beta^2-3)\sqrt{\frac{x}{3}}A^{(\Delta 1)(\ell)}=0\ ,
\\
&[x +\beta^2 \ell(\ell+1)] B^{(\Delta1)(\ell)}
+\frac{1}{2}(\beta^2-3)A^{(\Delta 1)(\ell)}=0\ ,
\end{align}
The determinant gives us four roots. The first is simple,
\begin{align}
x_1=-\beta^2 \ell(\ell+1) 
\quad\Rightarrow\quad
\Delta^{(1)}_{1\pm}=\frac32\pm\frac12\sqrt{1-4\beta^2\ell(\ell+1)}\ ,
\end{align}
and it is in the principal series for $\ell\geq1$. The other three are roots of the following cubic equation
\begin{align}
0=f(x):=&x^3 +[\beta^2(3\ell^2+3\ell+1)-9]x^2+[\beta^4(3\ell^4+6\ell^3+2\ell^2-\ell)-3\beta^2(3\ell^2+3\ell+2)+18]x
\nonumber\\
&+\beta^6 \ell^2(\ell+1)^2(\ell+2)(\ell-1)\ . 
\label{cubic}
\end{align}
Denoting the roots of this cubic equation by $x_i(\ell)$, the corresponding dS energies are
\be
\Delta_{1\pm}^{(i)}= \frac32 \pm \frac12 \sqrt{1+ 4x_i(\ell)}\ ,\qquad i=2,3,4\ .\label{d1pm}
\ee
We will suppress the argument $\ell$ below for simplicity in notation. These three roots have complicated expressions. We can show that they are real and non-positive as follows. First, we notice that when we evaluate $f(x)$ at $z=-\beta^2 \ell(\ell+1)$, we have $f(z)>0$ for $\ell\ge 1$. Next, we observe that $f(x)$ at $y=6-\beta^2 \ell(\ell+1)$ gives $f(y)\leq0$ for $\ell \ge 1$, and that $f(0)\ge 0$ again for $\ell\ge 1$. 
Noting that $z<y\leq0$, with the equality holding only when $\b^2=3$ and $\ell=1$. It follows that $f(x)$ must have two real roots $x_2$ and $x_3$, obeying $y<x_2\leq 0$ and $z<x_3<y$. Since the cubic equation has two real roots, the third root $x_4$ must be real, and $x_4<z$. 

For $\ell>1$, we can have a stronger condition
\begin{align}
\beta^2=\frac92\,:\qquad \ell>1:\qquad     x_4<z<x_3<y<x_2<-\frac{1}{4}\ ,
\end{align}
which means that for $\ell\geq2$, all modes are in the principal series. 
When $\ell=1$, we have
\begin{align}
 \beta^2=\frac92\,:\qquad \ell=1:\qquad  x_2=0\quad,\quad x_{3,4}=\frac{15}{4}(-3\pm 1) <-\frac14\ ,
\end{align}
which means that the conformal dimensions \eq{d1pm} for the roots $x_{3,4}$ are in the principal series and $x_2=0$ is in the exceptional series, and hence represents a massless mode. Note also that the masses are given by $m^2_i=-x_i/L^2$.

When $\b^2=3$, the expression in \eqref{cubic} simplifies coonsiderably, and the four modes of spin-1 sector are given by:
\begin{align}
\b^2=3: \quad \D^{(1)}_{1\pm}=\frac32\pm\frac12\sqrt{1-12(\ell-1)(\ell+2)} \ , \quad \D^{(i)}_{1\pm}=\frac32\pm\frac12\sqrt{1-12\ell(\ell+1)} \ ,\quad i=2,3,4 .
\end{align}
which implies that, for $\ell\geq2$, all modes are in the principal series. For $\ell=1$, the modes $\D^{(i)}_{1\pm}$ with $i=2,3,4$ are in the principal series, while $\D^{(1)}_{1\pm}=3/2\pm1/2$ is in the exceptional series and therefore represents a massless mode.

{\it The $\ell=0$ modes:} 

For $\ell=0$, this sector contains the fields $\left(\nabla^\mu h_{\mu\nu}^{(0)}, \nabla^\mu b_{\mu\nu}^{(0)},  a^{(0)}_\mu, b^{(0)}_{\mu\nu}\right) $. The first one contributes to the spin-0 sector by choosing the gauge choice 
\be
\nabla^\m h^{(0)}_{\m\n}=\frac12 \nabla_\n h^{(0)}\ ,
\label{tc}
\ee
and $\nabla^\mu b_{\mu\nu}^{(0)}$=0 from \eq{b33}. Thus, we are left with $a^{(0)}_\mu$ and $b^{(0)}_{\mu\nu}$. When $\b^2>3$, from \eq{b31}, and from the curl of \eq{a31}, defining $c_\mu := a^{(0)}_\mu +\frac{1}{2vk}\nabla_\mu b^{(0)}$, and using the spin-1 projector, we obtain 
\begin{align}
\b^2>3: & \quad \left(\Box_4 -\frac{4}{L^2}\right)\sqrt{\frac{\tv}{v}}b^{(0)T}_{\mu\nu}  + \frac12 \left(\frac{1}{a^2}-\frac{3}{L^2}\right)\frac{1}{k}*F_{\mu\nu}(c) = 0\ ,
\label{argued1}
\\
&\quad \left(\Box_4-\frac{5}{2L^2}-\frac{1}{2a^2}\right) \frac{1}{k}*F_{\mu\nu}(c) 
+ \left(\Box_4 -\frac{4}{L^2}\right)\sqrt{\frac{\tv}{v}}b^{(0)T}_{\mu\nu} = 0 \ ,
\label{argued2}
\end{align}
where $b^{(0)T}_{\mu\nu}$ and $c_\mu$ are divergent free, $F(c)=dc$ and we have used \eq{b33} and \eq{b34}. These equations can be diagonalized by defining $\Tilde{b}_{\mu\nu} := b^{(0)T}_{\mu\nu} + 3v *F_{\mu\nu}(c)$, thus obtaining
\begin{align}
\left(\Box_4-\frac{4}{L^2}\right) \Tilde{b}_{\mu\nu}=0\ \implies \ 
\nabla^\m \partial_{[\mu}\Tilde{b}_{\n\r]}=0\ ,
\end{align}
which means that $\Tilde{b}_{\m\n}$ is a massless 2-form field in dS$_4$, which is on-shell dual to a scalar field. Denoting this scalar by $\varphi$, we find 
\begin{align}
\partial_\m \varphi=\frac{1}{3!}\epsilon_{\m\n\r\s}\nabla^\n \Tilde{b}^{\r\s}\ \implies \ 
\epsilon_{\m\n\r\s}\nabla^\n b^{(0)T\r\s}=\partial_\m \varphi-6v\left(\Box_4-\frac{3}{L^2}\right)c_\m\ ,
\end{align}
where we have used the relation $\Tilde{b}_{\mu\nu} := b^{(0)T}_{\mu\nu} + 3v *F_{\mu\nu}(c)$ encountered above. 
Using this equation in \eqref{a31} for $\ell=0$, we obtain 
\begin{align}
\left(\Box_4-\frac{9}{2L^2}\right)c_\m +\frac{1}{6v}\partial_\m(\varphi+2b^{(0)})=0\ .
\label{spin1eq}
\end{align}
Projecting this equation to spin 1, we get $\left(\Box_4-\frac{9}{2L^2}\right)c_\m^T=0$ which yield the conformal dimension $\Delta=\frac32\pm i\frac{\sqrt{5}}{2}$, which is the root $x_4(\ell)$ evaluated at $\ell=0$ ( eee \eq{cubic} and the  subsequent discussion of the roots), while the projection to spin 0, does not give new information.

When $\b^2=3$, it follows from \eqref{b31} and \eqref{a31} that
\begin{align}
\b^2=3: & \qquad \left(\Box_4-\frac{4}{L^2}\right)b_{\m\n}=0\quad,\quad  \left(\Box_4 -\frac{3}{L^2}\right)a_\m=0\ .
\end{align}
The first equation describes a massless 2-form, which is dual to massless scalar, while the second equation describes a massless vector field. These correspond precisely to $\D^{(i)}_{1\pm}$ evaluated at $\ell=0$.

\subsubsection*{\boxed{\it Spin\, 0}}

{\it The $\ell \ge 1$ modes:} 

As displayed in \eq{exps}, we have the scalar field $\phi^{(\ell)}, N^{(\ell)}$ for $\ell\ge 0$, and $a_i^{(\ell)}$ for $\ell \ge 1$. The field $b^{(0)}$ is eaten $a_{\mu^{(0)}}$ as explained above. Using harmonic expansions on dS$_4$,  
\begin{align}
N^{(\ell)}(x) &=\sum_{\Delta,\ell\ge 0}^{(\Delta0)(\ell)}N^{(\D0)(\ell)}D^{(\Delta0)}(x)
\ ,\quad
a^{(\ell)}(x) =\sum_{\Delta,\ell\ge 1} a^{(\Delta0)(\ell)}D^{(\Delta0)}(x)\ ,
\nn\\
\phi^{(\ell)}(x) &=\sum_{\Delta,\ell\ge 0} \phi^{(\Delta0)(\ell)}D^{(\Delta0)}(x)\ .
\end{align}
The field equations give
\begin{align}
&\left[ \Delta(\Delta-3)-9 + \beta^2\ell(\ell+1)+\beta^2\right]L^2N^{(\Delta0)(\ell)} 
-3\beta^2 (\beta^2-3) \ell(\ell+1) a^{(\Delta0)(\ell)} = 0\ , 
\label{scalar1}
\\
&\left[\Delta(\Delta-3)+\beta^2\ell(\ell+1)\right]a^{(\Delta0)(\ell)}-L^2N^{(\Delta0)(\ell)} = 0\ ,\label{scalar2}
\\
&\left[ \Delta(\Delta-3)-6 + \beta^2 \ell(\ell+1) \right]\phi^{(\Delta0)(\ell)} = 0 \ ,
\label{scalar3}
\end{align}
For $\ell\ge 1$, the first two equations yield the result for the combinations of the fields $(N,a)$
\begin{align}
\Delta(\Delta-3)=&
\frac92-\frac12\beta^2[2\ell(\ell+1)+1] \pm \frac12\sqrt{81
-18\beta^2[2\ell(\ell+1)+1] + \beta^4 [12\ell(\ell+1)+1]}\ .
\end{align}
Thus, for the case of $\beta^2=9/2$ we have the following results for the combination of he fields $(N,a)$
\begin{align}
\beta^2=\frac92\,:\qquad & \Delta_{0\pm}^{(1)} = \frac{3}{2}\pm\frac{3}{2}\sqrt{1-2\ell(\ell+2)}\ ,
\nn\w2
& \Delta_{0\pm}^{(2)} = \frac{3}{2}\pm\frac{3}{2}\sqrt{3-2\ell^2}\ ,
\end{align}
implying the following unitary representations:
\begin{align}
\Delta_{0\pm}^{(1)} : \qquad &\mbox{Principal series:}\qquad \ell\ge 1\ ,
\\
\Delta_{0\pm}^{(2)} : \qquad &\mbox{Principal series:}\qquad \ell> 1
\ ,
\\
&\mbox{Exceptional series:} \quad \ell=1 \ .
\end{align}
The $\ell=1$ mode is massless. It is removed by the residual gauge symmetry \eqref{sss}.
Next, we consider the following case.
\begin{align}
\beta^2=3\,: \qquad & \Delta_{0\pm}^{(1)}=\frac32\pm\frac12\sqrt{9-12\ell(\ell+1)}\ . 
\nn\w2
& \Delta_{0\pm}^{(2)}=\frac32\pm\frac12\sqrt{9-12(\ell-1)(\ell+2)}\  ,
\end{align}
implying the following unitary representations:
\begin{align}
\Delta_{0\pm}^{(1)} : \qquad &\mbox{Principal series:}\qquad \ell\ge 1\ ,
\\
\Delta_{0\pm}^{(2)} : \qquad &\mbox{Principal series:}\qquad \ell> 1\ ,
\\
&\mbox{Exceptional series:} \quad \ell=1 \ .
\end{align}
The $\ell=1$ mode is massless. It is removed by the residual symmetry \eqref{sss}.

Finally, we turn to the equation for $\phi^{(\ell)}$ obeying \eq{scalar3} which readily gives 
\begin{align}
&  \Delta_{\phi\pm} =\frac32 \pm \frac12 \sqrt{33-4\b^2\ell(\ell+1)}\ .
\end{align}
It follows that
\begin{align}
\b^2=\frac92\,:\qquad\Delta_{\phi +}:\qquad & \mbox{Principal series:}\qquad 
\ell\geq1 \ ,
\\
\b^2=3\,:\qquad\Delta_{\phi +}:\qquad & \mbox{Principal series:}\qquad 
\ell>1 \ ,\nonumber
\\
& \mbox{Exceptional series:}\quad 
\ell=1 \ .
\end{align}

\subsection*{\it The $\ell=0$ modes:}

For $\ell=0$ in the spin-0 sector, we have the fields $(\nabla^\mu\nabla^\nu h^{(0)}_{\mu\nu},h^{(0)\mu}_\mu,\nabla^\mu\nabla^\nu b^{(0)}_{\mu\nu},  \nabla^\mu a^{(0)}_\mu, b^{(0)}, N^{(0)}, \phi^{(0)}) $. 

From \eq{b33} and the gauge choice in \eq{ctg}, we have $\nabla^\mu\nabla^\nu h^{(0)}_{\mu\nu}$ is determined by $h^{(0)\mu}_\mu$, which in turn is related to $N^{(0)}$ from the
trace of \eq{en31}, which upon using \eq{tc} gives
\be
h^{(0)\mu}_\mu=L^2\left(\Box_4 N^{(0)}+4k^2 f N^{(0)}\right)\ .
\ee
The scalar $\nabla^\mu \nabla^\nu b^{(0)}_{\mu\nu}=0$, due to from \eq{b33}. When $\b^2>3$, $\nabla^\mu a_\mu^{(0)}$ and $b^{(0)}$ they always arise in the combination $c_\mu = a^{(0)}_\mu +\frac{1}{2vk}\nabla_\mu b^{(0)}$, and given that the scalar $\nabla^\mu c_\mu=0$, we are left with $N^{(0)}$ and $\phi^{(0)}$. We find from \eq{en34} and \eq{d31} they obey 
\begin{align}
\beta^2=\frac92\,:\qquad \left(\Box_4+\frac{9}{2L^2}\right)N^{(0)}=0\ ,
\ \left(\Box_4+\frac{6}{L^2}\right)\phi^{(0)}=0 \ .
\label{d1}
\end{align}
We see that both scalars are tachyonic. 

When $\b^2=3$, the analysis is the similar to the pervious case, one finds three physical modes, namely $(b^{(0)},N^{(0)},\phi^{(0)})$, from \eqref{urs}, \eqref{en34} and \eqref{d31}, one obtains
\begin{align}
\beta^2=3\,:\qquad \Box_4b^{(0)}=0\ , 
\ \left(\Box_4+\frac{6}{L^2}\right)N^{(0)}=0 \ ,\ 
\left(\Box_4+\frac{6}{L^2}\right)\phi^{(0)}=0\ ,
\label{d20}
\end{align}
We see that there is one massless scalar and there are two tachyons in this case. 

\subsection{The fermionic sector}


\subsubsection*{\boxed{\it Spin\, 3/2}}
The harmonic expansion in $dS_4$ for $\ell \ge 1$ is given by
\begin{align}
\psi^{(\ell)}_{\mu\pm}(x)=\sum_{\Delta} \psi^{(\Delta,\frac32)(\ell)}_\pm D^{(\Delta,\frac32)}_{\mu\pm}(x)
+\sum_{\Delta} \psi^{(\Delta,\frac12)(\ell)}_\pm D^{(\Delta,\frac12)}_{\mu\pm}(x)\ ,
\label{fhe}
\end{align}
where $\pm$ represent the chiral projections, $\psi_\pm$ are constant expansion coefficients, and the tensor spinor indices are carried by the representation functions $D_{\mu\pm}(x)$. 
From \eq{3/2+1} and \eq{3/2+2}, projecting to spin-3/2 sector and for $\ell \ge 1$ we obtain
\begin{align}
&\frac{1}{L}\left(-\Delta+\frac{3}{2}\right)\psi^{(\Delta,\frac{3}{2})(\ell)}_+
-iM_\ell\psi^{(\Delta,\frac{3}{2})(\ell)}_- = 0 \ ,
\\
&\frac{1}{L}\left(-\Delta+\frac{3}{2}\right)\psi^{(\Delta,\frac{3}{2})(\ell)}_-
-iM_\ell\psi^{(\Delta,\frac{3}{2})(\ell)}_+ = 0\ ,
\end{align}
which admits solution provided that 
\begin{align}
\D_{\frac32\pm}=\frac{3}{2}\pm \sqrt{-\b^2\ell(\ell+1)+\frac{3}{2}-\frac{27}{4\b^2}}\ ,
\end{align}
which belongs to the principal series according to \eq{A3/2} for both values $\b^2=3$ and $\b^2=9/2$.

For $\ell=0$ in the spin-3/2 sector, the residual symmetry \eq{rsf1} suggests that $\lambda^{(0)}_+$ is eaten by $\psi^{(0)}_{\mu+}$. In more detail, from \eqref{fz1}, \eqref{fz2} and \eqref{fz4}, we obtain
\begin{align}
&\slashed{\nabla} \Tilde{\psi}_\mu-\frac12 \nabla_\mu \gamma^\nu \Tilde{\psi}_\nu=0\ , \label{gt0}
\\
&\slashed{\nabla}\Psi^{(0)}_-=0\ ,\label{gt1}
\\
&\gamma^\mu\Tilde{\psi}_\mu =-4\Psi^{(0)}_-\ ,\label{gt2}
\end{align}
where we have defined 
\begin{align}
    \Tilde{\psi}_\mu:=\psi^{(0)}_{\mu+}+3if\nabla_\mu \lambda^{(0)}_+\ ,
\end{align}
which is invariant under the residual symmetry. By taking the gamma trace of \eqref{gt0}, we derive $\nabla^\mu \Tilde{\psi}_\mu=0$. Thus, together with \eqref{gt1} and \eqref{gt2}, we can express gravitino as $
\Tilde{\psi}_\mu=\Tilde{\psi}^{(\Delta,3/2)}D^{(\Delta,3/2)}_{\mu+}(x)-\gamma_\mu \Psi^{(0)}_-$. Plugging this expression  in \eqref{gt0}, we have
\begin{align}
\frac{1}{L}\left(\Delta-\frac{3}{2}\right)\Tilde{\psi}^{(\Delta,\frac{3}{2})}=0\ ,
\end{align}
which gives $\Delta=3/2$, implying that $\Tilde{\psi}_{\mu}$ belongs to the discrete series, in accordance with \eqref{h2f}.

\subsubsection*{\boxed{\it Spin\, 1/2}}

The modes in the spin-1/2 sector are
($\nabla^\mu\psi^{(\ell)}_{\mu\pm},\ \gamma^\mu\psi^{(\ell)}_{\mu\pm},\ \Psi^{(\ell)}_\pm, \chi^{(\ell)}_\pm,\ \lambda^{(\ell)}_\pm$) for $\ell \ge 1$. We find that $\nabla^\mu\psi^{(\ell)}_{\mu\pm}$ and $\gamma^\mu\psi^{(\ell)}_{\mu\pm}$ are not physical modes. The dilatini $\chi_\pm^{(\ell)}$  decouple from the other fields due to $f'=0$, and from \eq{dilatino11} and \eq{dilatino2} they obey
\begin{align}
&\frac{1}{L}\left(\Delta-\frac{3}{2}\right)\chi^{(\Delta,\frac{1}{2})(\ell)}_++iM_\ell\chi^{(\Delta,\frac{1}{2})(\ell)}_- = 0 \ ,
\\
&\frac{1}{L}\left(\Delta-\frac{3}{2}\right)\chi^{(\Delta,\frac{1}{2})(\ell)}_-+iM_\ell\chi^{(\Delta,\frac{1}{2})(\ell)}_+ = 0 \ .
\end{align}
Solving these equations requires that 
\begin{align}
\D_{\frac12\pm}=\frac{3}{2}\pm \sqrt{-\b^2\ell(\ell+1)+\frac{3}{2}-\frac{27}{4\b^2}}\ ,
\end{align}
which shows that there are two towers massive spin 1/2 fields in the principal series representation for $\ell \ge 1$.  

The other four fields are ($\Psi^{(\ell)}_\pm, \lambda^{(\ell)}_\pm$) and the four associated equations are
\begin{align}
&\frac{i}{L}\left(\Delta-\frac{3}{2}\right)\Psi^{(\Delta,\frac{1}{2})(\ell)}_+ - M_\ell\Psi ^{(\Delta,\frac{1}{2})(\ell)} _- + \frac{1}{2}(3kf+1)\lambda ^{(\Delta,\frac{1}{2})(\ell)} _- = 0 \ ,
\\
&\frac{i}{L}\left(\Delta-\frac{3}{2}\right)\Psi ^{(\Delta,\frac{1}{2})(\ell)} _--M_\ell\Psi ^{(\Delta,\frac{1}{2})(\ell)} _+-\frac{1}{2}(3kf-1)\lambda ^{(\Delta,\frac{1}{2})(\ell)} _+ = 0 \ ,
\\
&\frac{i}{L}\left(\Delta-\frac{3}{2}\right)\lambda ^{(\Delta,\frac{1}{2})(\ell)} _++M_\ell\lambda ^{(\Delta,\frac{1}{2})(\ell)} _--\frac{1}{f}(kf+1)\Psi ^{(\Delta,\frac{1}{2})(\ell)} _- = 0 \ ,
\\
&\frac{i}{L}\left(\Delta-\frac{3}{2}\right)\lambda ^{(\Delta,\frac{1}{2})(\ell)} _-+M_\ell\lambda ^{(\Delta,\frac{1}{2})(\ell)} _++\frac{1}{f}(kf-1)\Psi ^{(\Delta,\frac{1}{2})(\ell)} _+ = 0 \ ,
\end{align}
again for $\ell \ge 1$. Performing the harmonic expansions on dS$_4$ and requiring that the determinant of the coefficient matrix vanishes for the solutions to exist gives the following result:
\begin{align}
\D^{(1)}_{\frac12\pm}&=\frac32\pm\sqrt{-\frac12\b^2+\frac92 -L^2 M^2_{\ell\b}+\frac12\sqrt{\b^2-3}\sqrt{\b^2+9+12L^2M^2_{\ell\b}}}\ ,
\label{spinhalf1}
\\
\D^{(2)}_{\frac12\pm}&=\frac32\pm\sqrt{-\frac12\b^2+\frac92 -L^2 M^2_{\ell\b}-\frac{1}{2}\sqrt{\b^2-3}\sqrt{\b^2+9+12L^2M^2_{\ell\b}}}\ ,
\label{spinhalf2}
\end{align}
where 
\begin{align}
L^2 M^2_{\ell\b}=\b^2\ell(\ell+1)-\frac{3}{2}+\frac{27}{4\b^2}\ .
\end{align}
Thus we have
\begin{align}
\b^2=\frac92:  \qquad \Delta^{(1)}_{\frac12\pm}=\frac{3}{2}\pm\frac{3}{2}\sqrt{-2\ell^2+2}\ , \qquad \Delta^{(2)}_{\frac12\pm}=\frac{3}{2}\pm\frac{3}{2}\sqrt{-2\ell(\ell+2)}\ ,
\end{align}
When $\ell=1$, $\Delta^{(1)}_{\frac12\pm} = \frac{3}{2}$, which means that the fields associated with this $\Delta$ are  massless. Altogether we have 
\begin{align}
\Delta_{\frac12\pm}^{(1)} : \qquad &\mbox{Principal series:}\qquad \ell\ge2\ ,
\\
&\mbox{Discrete series:}\qquad \ \ \ell=1\ ,
\\
\Delta_{\frac12\pm}^{(2)} : \qquad &\mbox{Principal series:}\qquad \ell\ge1\ .
\end{align}

For $\ell=0$ in the spin 1/2 sector, the independent fields are ($\chi^{(0)}_-$, $\Psi^{(0)}_-$). From \eqref{fz3} and \eqref{gt1}
\begin{align}
\frac{1}{L}\left(\Delta-\frac{3}{2}\right)\chi^{(\Delta,\frac{1}{2})(0)}_-=0\ ,\ 
\frac{1}{L}\left(\Delta-\frac{3}{2}\right)\Psi^{(\Delta,\frac{1}{2})(0)}_-=0\ .
\end{align}
It follows that $\Delta_{\frac12}=\frac{3}{2}$ for both fields\footnote{In the case of Mink$_4\times S^2$ spectrum, only one massless state arises in this sector.}. Thus,  $\chi^{(0)}_-$ and $\Psi^{(0)}_-$ are in the discrete series representations and massless. 

Next, from \eq{spinhalf1}  and \eq{spinhalf2} , we get
\begin{align}
\b^2=3: \qquad \Delta^{(1)}_{\frac12\pm}=\Delta^{(2)}_{\frac12\pm}=\frac32\pm\sqrt{3-3\left(\ell+\frac12\right)^2}\ ,
\end{align}
from which it follows that
\begin{align}
&\Delta_{\frac12\pm}^{(1)}\mbox{ and } \Delta_{\frac12\pm}^{(2)} : \qquad \mbox{Principal series:}\qquad \ell\ge\frac32\ .
\end{align}
In the $\ell=\frac12$ sector, the residual symmetry \eqref{rsf3} suggests that $\l^{(1/2)}_\pm$ is eaten by $\psi^{(1/2)}_{\m\pm}$. Using \eqref{3/2+1}-\eqref{3/2+4}, \eqref{gaugino11} and \eqref{gaugino2}, we obtain 
\begin{align}
\slashed{\nabla}\Tilde{\psi}_{\m\pm} \pm i\frac1a \Tilde{\psi}_{\m\mp}=0 \ ,\  
\slashed{\nabla}\Tilde{\Psi}_\pm=0 \ ,\ 
\gamma^\m \Tilde{\psi}_{\m\pm}\pm2\Tilde{\Psi}_\mp=0 \ ,
\end{align}
where $\Tilde{\psi}_{\m\pm}:=\psi^{(1/2)}_{\mu\pm}+ia^2\nabla_\mu \lambda^{(1/2)}_\pm$ and $\Tilde{\Psi}_\pm :=\Psi^{(1/2)}_\pm-\frac{a}{2}\nabla_\m \l^{(1/2)}_\pm$ which are invariant under the residual symmetry. The above equations also imply that $\nabla^\m\Tilde{\psi}_{\m\pm}=-\frac{i}{a}\Tilde{\Psi}_\mp$. Hence there are two independent fields $\Tilde{\Psi}_\pm$, and the field equations imply that they are in the discrete series representations and massless.

\subsection{Summary}

The complete spectrum for the $dS_4\times S^2$ vacuum solution is summarized in Table 2 and Table 3 for the cases of $\beta^2=9/2$ and $\b^2=3$, respectively. Both spectra have no supersymmetry. In the first case, one spin-2, two spin-$3/2$, four spin-1, six spin-$1/2$ and three spin-0 towers of states arise that are massive in the de Sitter sense. Two of the spin-0 towers have single real tachyonic modes at their lowest level, one coming from the volume of the internal space and the other from the dilaton. The massless sector consists of one spin-2, three spin-1, five spin-$1/2$ and one spin-0 states. 
\begin{table}[H]
\centering
\begin{tabular}{|c|c|c|}
\hline
 s & $\Delta$ & \mbox{Comments}\\
\hline\hline
$2$ & $\frac32 + \frac32\sqrt{1-2\ell(\ell+1)}$ & $\ell\ge 0$, \ \mbox{massless for}\  $\ell=0$
\\
\hline
$\frac32$  & $\frac{3}{2}+ \frac{3}{2}\sqrt{-2\ell(\ell+1)}$ & 
$\ell\geq0$\mbox{, complex massive gravitino for}\ $\ell\ge1$
\\
&&
\mbox{massive for}\ $\ell=0$ \mbox{(discrete series)}\\
\hline
& $\frac{3}{2}+\frac{3}{2}\sqrt{\frac{1}{9}-2\ell(\ell+1)}$ & $\ell\geq1$, \
\\
$1$  & $\frac32+\frac12 \sqrt{1+4x_i(\ell)}$, $i=2,3,4$ & $\ell\geq1$, $x_i(\ell)$ are roots of \eq{cubic}  
\\
&&\mbox{ $x_2=0$ massless for} $\ell=1$
\\
&&\mbox{ $x_3=0$ for} $\ell=0$ (massless 2-form)
\\
& $\frac32+i\frac{\sqrt{5}}{2}$ & \mbox{ $x_4=-\frac32 $ for} $\ell=0$ (massive vector)
\\
\hline
&  $\frac{3}{2}+ \frac{3}{2}\sqrt{-2\ell(\ell+1)}$& $\ell\geq0$ \mbox{, complex massive spinor for} $\ell\ge1$
\\
&&\mbox{massless for} $\ell=0$
\\
$\frac12$&  $\frac{3}{2}+ \frac{3}{2}\sqrt{-2\ell^2+2}$ & $\ell\geq1$
\mbox{, complex massive spinor for}\ $\ell\geq2$ \\
&&\mbox{massless for}\ $\ell=1$
\\
&  $\frac{3}{2}+ \frac{3}{2}\sqrt{-2\ell(\ell+2)}$& $\ell\geq0$
\mbox{, complex massive spinor for}\ $\ell\ge1$
\\
&&\mbox{{massless} for}\ $\ell=0$
\\
\hline
& $\frac{3}{2}+ \frac{3}{2}\sqrt{1-2\ell(\ell+2)}$& 
$\ell\ge 1$
\\
$0$ & $\frac{3}{2}+ \frac{3}{2}\sqrt{3-2\ell^2}$ & 
$\ell\ge 0$, \mbox{{\bf tachyon} for}\ $\ell=0$
\\
& $\frac{3}{2}+ \frac{3}{2}\sqrt{\frac{11}{3}-2\ell(\ell+1)}$ & $\ell\ge 0$, \mbox{{\bf tachyon} for}\ $\ell=0$
\\
\hline
\end{tabular}
\caption{The full spectrum on dS$_4\times S^2$ with nonvanishing $U(1)_{R+}$ flux. The $SO(4,1)$ representations with lowest weight $(\Delta,s)$ that arise are listed. The massless sector consist of a graviton, three vectors, five spin-1/2  fermions and a scalar which arises from the dualization of a massless 2-form. The massive KK modes consist of a spin-2 tower, two spin 3/2 towers, four spin-1 towers, six spin-1/2 towers and three spin-0 towers, two of which harbor single tachyonic states at the bottom. One triplet and one singlet massless spin-1/2 states are removed by the residual shift symmetry \eqref{rsf2} and \eqref{rsf1}, respectively. A triplet of massless spin-0 states are removed by the residual shift symmetry \eqref{sss}.}
\end{table}
\begin{table}[H]
\centering
\begin{tabular}{|c|c|c|}
\hline
 s & $\Delta$ & \mbox{Comments}\\
\hline\hline
$2$ & $\frac32 + \frac12\sqrt{9-12\ell(\ell+1)}$ & $\ell\ge 0$, \ \mbox{massless for}\  $\ell=0$
\\
\hline
$\frac32$  & $\frac{3}{2}+ i\sqrt{3}\left(\ell+\frac12\right)$ & 
\mbox{$\ell\geq1/2$, complex massive gravitino for}\ $\ell\ge1/2$
\\
\hline
& $\frac32+\frac12\sqrt{1-12(\ell-1)(\ell+2)}$ & $\ell\geq1$, massless for $\ell=1$\
\\
$1$& $\frac32+\frac12\sqrt{1-12\ell(\ell+1)}$ & $\ell\geq1$ \
\\
& $\frac32+\frac12\sqrt{1-12\ell(\ell+1)}$ & $\ell\geq0$, massless 2-form for $\ell=0$\
\\
& $\frac32+\frac12\sqrt{1-12\ell(\ell+1)}$ & $\ell\geq0$, massless vector for $\ell=0$\
\\
\hline
&  $\frac{3}{2}+ i\sqrt{3\left(\ell+\frac12\right)\left(\ell-\frac12\right)}$ & 
\mbox{$\ell\geq3/2$, complex massive spinor for}\ $\ell\geq3/2$ \\
$\frac12$&  $\frac{3}{2}+ i\sqrt{3\left(\ell+\frac12\right)\left(\ell-\frac12\right)}$& 
\mbox{$\ell\geq1/2$, complex massive spinor for}\ $\ell\ge3/2$
\\
&&\mbox{complex massless for}\ $\ell=1/2$
\\
&  $\frac{3}{2}+ i\sqrt{3}\left(\ell+\frac12\right)$& \mbox{$\ell\geq1/2$, complex massive spinor for} $\ell\ge1/2$
\\
\hline
& $\frac32+\frac12\sqrt{9-12\ell(\ell+1)}$&$\ell\geq0$, massless for $\ell=0$
\\
$0$ & $\frac32+ \frac12\sqrt{9-12(\ell-1)(\ell+2)}$ & 
$\ell\ge 0$, \mbox{{\bf tachyon} for}\ $\ell=0$
\\
& $\frac32+ \frac12\sqrt{33-12\ell(\ell+1)}$ & $\ell\ge 0$, \mbox{{\bf tachyon} for}\ $\ell=0$
\\
\hline
\end{tabular}
\caption{The spectrum on dS$_4\times S^2$ with vanishing $U(1)_{R+}$ flux. The $SO(4,1)$ representations with lowest weight $(\D,s)$ are listed. The massless sector consist of a graviton, four vectors(one triplet and one singlet), four spin-1/2 fermions(two doublets) and two scalars in which one arises form the dualization of a massless 2-form. The massive KK modes consist of a spin-2 tower, two spin-3/2 towers, four spin-1 vector towers, six spin-1/2 towers and three spin-0 towers, two of which harbor single tachyonic states at the bottom. Two doublet massless spin-1/2 states are removed by the residual shift symmetry \eqref{rsf3}. A triplet of massless spin-0 states are removed by the residual shift symmetry \eqref{sss}. In this table, $\ell=0,1,\dots$ for bosonic fields, and $\ell=\frac12,\frac32,\dots$ for fermionic fields.}
\end{table}

\section{Flow from dS\texorpdfstring{$_4\times S^2$}{4 x S2} to Mink\texorpdfstring{$_4\times S^2$}{4 x S2}}

In the previous section, we have seen that for the $dS_4\times S^2$ solution with $\beta^2=9/2$ there exist two tachyonic modes at the lowest Kaluza-Klein level. It is therefore natural to ask whether small perturbations along these directions drive the solution towards a stable vacuum. In this section, we show that turning on tiny perturbations of these tachyonic modes around the $\beta^2=9/2$ vacuum at $t=0$ indeed triggers a flow asymptotic to $Mink_4\times S^2$ as $t\to\infty$. We shall not carry out a similar analysis for the case of $\beta^2=3$, because the corresponding effective potential has a runaway behavior, as it is not bounded from below and decreases towards $-\infty$ along the unstable direction. This makes the solution physically unattractive, in contrast with the $\beta^2=9/2$ case, which we examine in detail below. We start with the following ansatz:
\begin{align}
&ds^2_6 = - e^{2n(t)} dt^2+ e^{2a(t)} [ (dx^1)^2 +(dx^2)^2 +(dx^3)^2 ] +e^{2b(t)} (d\theta^2 +\sin^2 \theta d\varphi^2)\ , \nonumber
\\
&\quad F_{(2)}=\frac{1}{2}\sin\theta d\theta\wedge d\varphi\ , \quad
\phi=\phi(t)\ .
\end{align}
Plugging the ansatz above into \eqref{Lagrangian}, we obtain
the effective action 
\begin{align}
S=2\pi V_3\int dt e^{-n+3a+2b}\left(-3\dot{a}^2-6\dot{a}\dot{b}-\dot{b}^2+\frac12\dot{\phi}^2\right)+e^{n+3a+2b}\left(e^{-2b}-\frac{1}{4}f e^{-4b}-\frac{1}{f}\right)\ ,
\end{align}
which is invariant under reparameterization of time, i.e. $t\rightarrow t'(t)$. By varying $n$ , we get the constraint equation:
\begin{align}
\delta n:\qquad 3\dot{a}^2+6\dot{a}\dot{b}+\dot{b}^2-\frac12 \dot{\phi}^2+e^{-2b}-\frac{1}{4}fe^{-4b}-\frac{1}{f}=0\ .
\label{cc}
\end{align}
Now, by redefining time, we can effectively set $n=0$, and the resulting 1D Lagrangian is given as
\begin{align}
\mathcal{L}= e^{3a+2b}\left(-3\dot{a}^2-6\dot{a}\dot{b}-\dot{b}^2+\frac12 \dot{\phi}^2+e^{-2b}-\frac14 fe^{-4b}-\frac{1}{f}\right)\ .
\end{align}
The equations of motion for $a,\, b$ and $\phi$ are 
\begin{align}
\delta\phi:&\qquad \ddot{\phi}+(3\dot{a}+2\dot{b})\dot{\phi}+\frac14 f' e^{-4b}-\frac{f'}{f^2}=0\ ,
\\
\delta a:&\qquad 2\ddot{a}+2\ddot{b}+4\dot{a}\dot{b}+3\dot{b}^2+3\dot{a}^2+\frac12\dot{\phi}^2+e^{-2b}-\frac14fe^{-4b}-\frac{1}{f}=0\ ,
\\
\delta b:&\qquad 3\ddot{a}+\ddot{b}+6\dot{a}^2+3\dot{a}\dot{b}+\dot{b}^2+\frac12\dot{\phi}^2+\frac14fe^{-4b}-\frac1f=0\ ,
\end{align}
where due to the constraint equation \eqref{cc}, only two of the three equations are independent. 
We introduce new variables as follows
\begin{align}
t:={\sqrt{2}}(v\Tilde{v})^{\frac{1}{4}}\t\ ,\ \varphi:=\phi+\frac{1}{2}\ln\left(\frac{v}{\Tilde{v}}\right)\ ,\ 
B:= b+\frac12 \ln2+\frac14 \ln(v\tv)\ ,\ A:=a\ .
\end{align}
Then we have
\begin{align}
f=2\sqrt{v\Tilde{v}}\cosh\varphi \ ,\ 
f'=2\sqrt{v\Tilde{v}}\sinh\varphi\ ,
\end{align}
where and below prime denotes derivative with respect to $\t$.
In terms of the new variables, constraint equation, and dynamic equations become
\begin{align}
&3A'^2+6A'B'+B'^2-\frac12 \varphi'^2+e^{-2B}-\frac{1}{4} e^{-4B}\cosh\varphi-\frac{1}{\cosh\varphi}=0\ ,
\\
&\varphi''+(3A'+2B')\varphi'+\sinh\varphi\left(\frac14 e^{-4B}-\frac{1}{\cosh^2\varphi}\right)=0\ ,
\\
&A''+(3A'+2B')A'+\frac18e^{-4B}\cosh\varphi-\frac{1}{2\cosh\varphi}=0\ ,
\\
&B''+(3A'+2B')B'+ e^{-2B}-\frac38 e^{-4B}\cosh\varphi-\frac{1}{2\cosh\varphi}=0\ ,
\end{align}
where we have also taken combinations of the original equations for simplification. 
Setting $B'=B''=\varphi'=\varphi''=0$, we find the following solutions
\begin{align}
{\rm dS}_4\times S^2:\quad &\varphi=0\ ,\quad \ B=\frac12\ln\frac32\ , \quad A'=\pm\frac{2}{3\sqrt{3}}\ ,\ A''=0\ ,
\\
{\rm Mink}_4\times S^2:\quad &2e^{2B}=\cosh\varphi\ ,\quad \mbox{$B$ and $\varphi$ are constants}\ ,\quad A'=0\ ,\ A''=0\ .\ 
\end{align}
We give one numerical example in Figs.\eq{fig:phi}-\eq{fig:mink}. We show two slices of the phase space in Fig. \eq{fig:phi} and Fig. \eq{fig:b}. The first slice is given in coordinate ($\varphi,\varphi'$), and the second in ($B,B'$). We can see that the Minkowski solution is an attractor. The $dS_4$ solution corresponds to $(0,0)$ and $(\frac12\ln\frac32,0)$. We set the initial value at $\tau=0$ which is slightly away from dS vacuum
\be
\varphi=10^{-4},\quad \varphi'=0\ , \quad
B=\frac12\ln\frac32-2\times 10^{-3}\ ,\quad B'=0\ .
\ee
In Fig \eq{fig:mink}, the time-evolutions of $\varphi$ and $b$ are exhibited. One can see that after several damping oscillations, the two variables approach constants, and the combination $(2e^{2B}-\cosh\varphi)$ approaches zero, which agrees with the Mink$_4\times S^2$ solution.

\begin{figure}
  \centering
  \begin{minipage}{0.49\linewidth}
    \centering
    \includegraphics[width=\linewidth]{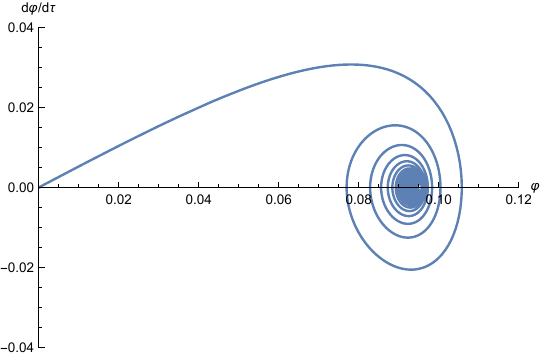}
    \caption{The phase-space diagram of ($\varphi,\varphi'$), the initial point is $(10^{-4},0)$. Here, $\tau\in(0,150)$.}
    \label{fig:phi}
  \end{minipage}\hfill
  \begin{minipage}{0.49\linewidth}
    \centering
    \includegraphics[width=\linewidth]{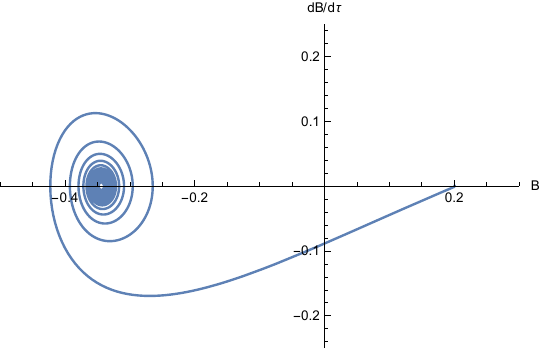}
    \caption{The phase-space diagram of ($B,B'$), the initial point is $(\frac12\ln\frac32-2\times10^{-3},0)$.}
    \label{fig:b}
  \end{minipage}
\end{figure}
\begin{figure}[H]
\centering
\includegraphics[width=0.6\textwidth]{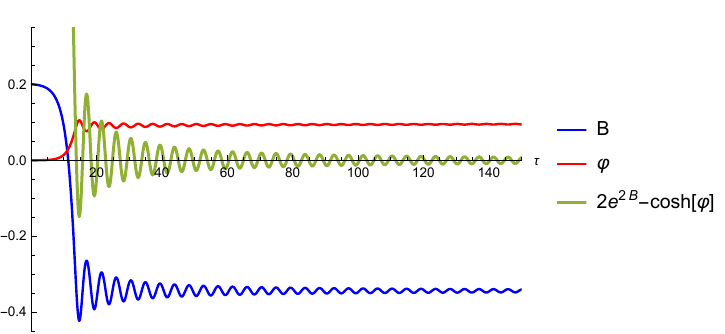}
\caption{$\varphi$ and $B$ approach constant quickly after several damping oscillations, and $(2e^{2B}-\cosh{\varphi})$ approaches zero.}
\label{fig:mink}
\end{figure}

\section{Conclusion and discussion}

We have investigated the consequence of including the Green-Schwarz anomaly counterterm and the associated terms required by supersymmetry in R-symmetry gauged $N=(1,0), 6D$ supergravity in determining its vacuum solutions. The solutions we have found are summarized in Table \ref{tab:solutions}. In all these solutions, the direct contribution of the GS counterterm which violates the gauge invariance vanishes, but the terms associated by supersymmetry are gauge invariant and they are essential for the solutions. The bosonic part of the Lagrangian is displayed in \eq{Lagrangian}, \eq{mod1} and \eq{mod2}, and the vacuum solutions are summarized in Table 4. 

\begin{table}[H]
\centering
\begin{tabular}{lll}
\hline
Solution  & $U(1)_{R+}$ flux ($k$) & $U(1)'$ flux (c) \\
\hline
$\mathrm{dS}_2 \times S^4$   & $0 \le k^2 < (12v\tilde v)^{-1}$ & 0  \\
$\mathrm{Mink}_2 \times S^4$ & $k^2 = (12v\tilde v)^{-1}$  & 0   \\
$\mathrm{AdS}_2 \times S^4$  & $k^2 > (12v\tilde v)^{-1}$  & 0    \\
$\mathrm{dS}_4 \times S^2$  & $k = 0$ & 0   \\
$\mathrm{dS}_4 \times S^2$  & $k^2 = (36v\tilde v)^{-1}$ & 0   \\
$\mathrm{dS}_4 \times S^2$  & 0  & $c \neq0$\\
$\mathrm{Mink}_4 \times S^2$ & $k^2 = (ve^{\phi_0}+\tilde v e^{-\phi_0})^{-1}$ & 0 \\
$\mathrm{AdS}_4 \times S^2$  & 0 & $ c \neq 0$  \\
$\mathrm{dS}_6$              & 0 & 0   \\
\hline
\end{tabular}
\caption{Summary of the vacuum solutions in this paper, indicating the supporting fluxes. The $U(1)_R$ flux is turned on $S^2$, the $U(1)'$ flux on (A)dS$_2$. The flux parameter $c$ is given in \eq{ddp}. In the first five solutions, and the dS$_6$ case, $e^{2\phi_0}=\tv/v$, which requires that $v\tv >0$. The solutions with $c\ne 0$ exist for nonvanishing $\tv$ and/or $\tv_1$. Only the Mink$_4\times S^2$ vacuum has supersymmetry, and it is half-maximal. Diagonal gauging is not required only for $\mathrm{Mink}_4 \times S^2$. The existence of the $\mathrm{(A)dS}_4 \times S^2$ solutions requires that the quartic equation \eq{qep}, which depends of the anomaly coefficients, admits at least one positive root. We have found one such model with the anomaly polynomial \eq{m2}.}
\label{tab:solutions}
\end{table}

The dS$_4 \times S^2$ solution requires a certain sign for the gauge anomaly. If we insist on the R-symmetry group to be the $U(1)_R$, under which the hyperfermions are neutral, the desired sign does not arise. However, considering an admixture of an external $U(1)$ such that the gauge group is a diagonal subgroup of $U(1)_R \times U(1)$, denoted by $U(1)_{+R}$, under which the hyperfermions are now charged, as was done in \cite{Suzuki:2005vu}, we find that models exist in which the gauge anomaly has the desired sign. We have tabulated a number of such model in Appendix A. It is remarkable that diagonal gauging is needed in order to have the 4D de Sitter compactification. This motivates further studies of such gaugings.

\begin{figure}
  \centering
  \begin{minipage}{0.48\linewidth}
    \centering
    \includegraphics[width=\linewidth]{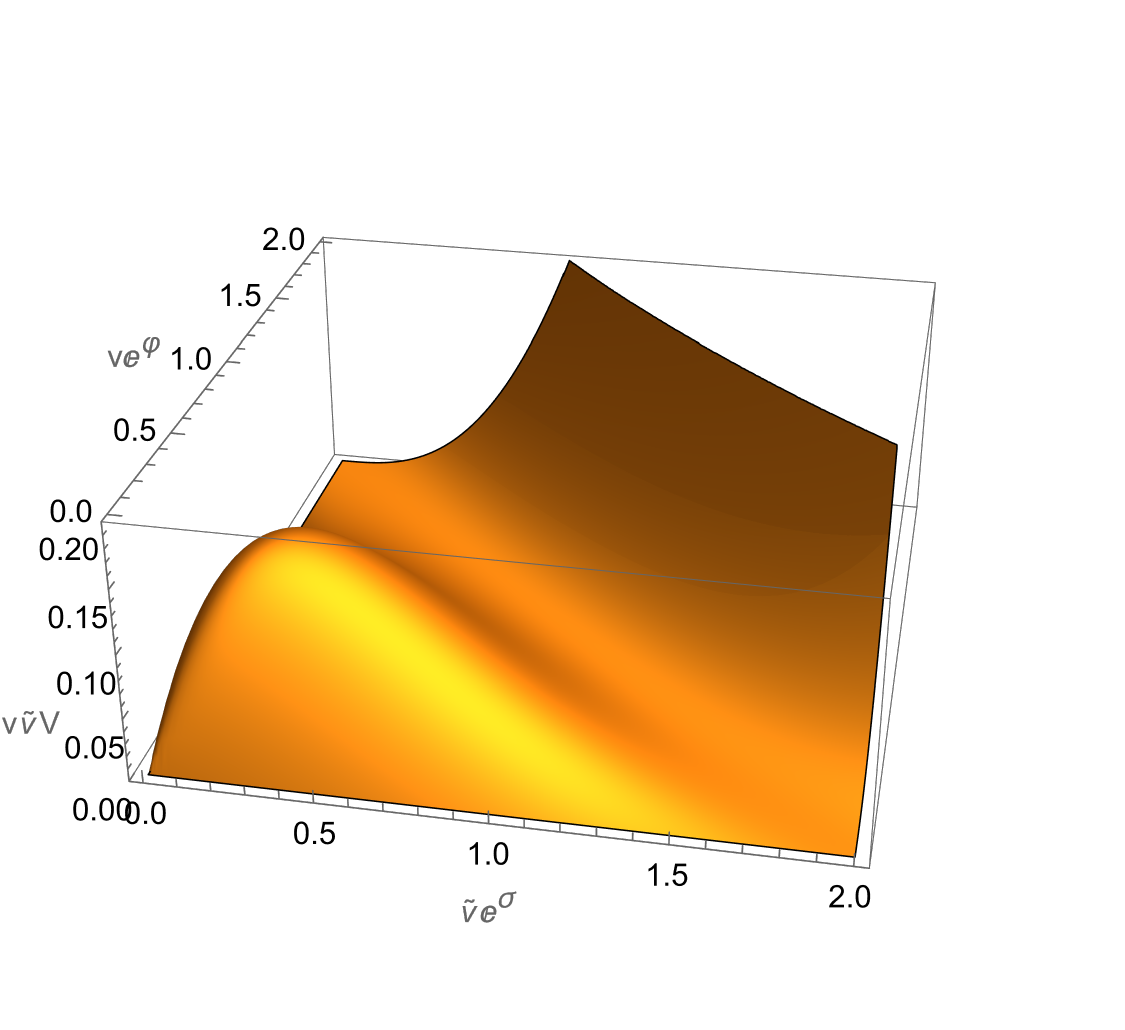}
    \caption{Scalar potential in 4D. The local maximum corresponds to the de Sitter, and the $ve^{\varphi}+\tv e^{\sigma}=2$ valley to the Minkowski vacua. }
    \label{fig:left}
  \end{minipage}\hfill
  \begin{minipage}{0.48\linewidth}
    \centering
    \includegraphics[width=\linewidth]{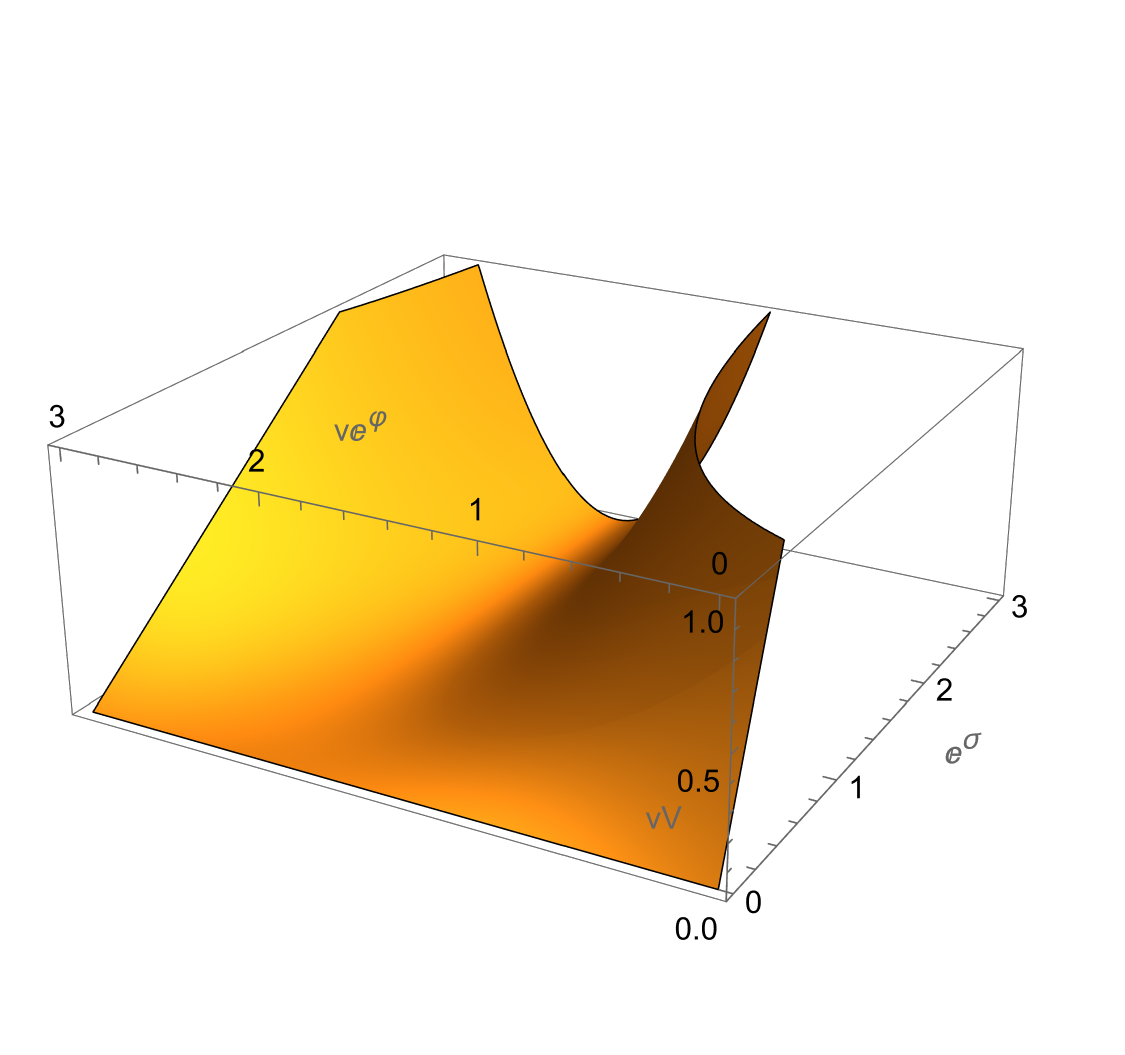}
    \caption{The scalar potential in 4D for $\tv=0$. The valley at $ve^{\varphi}=2$ denotes the Minkowski vacuum.}
    \label{fig:right}
  \end{minipage}
\end{figure}

Next, we computed the complete spectra for the theory compactified on Mink$_4\times S^2$ and dS$_4\times S^2$. In the first case, the masses of spin 0,1/2,1 KK towers get modified by a parameter proportional to the gauge anomaly. The full spectrum is unitary. In the absence of the anomaly,  the spectrum reduces to the one whose massless sector was known, while the massive sector was stated in \cite{Salam:1984cj} without a derivation. In the case of dS$_4\times S^2$, we find that at the bottom of two of the spin-0 KK towers there are tachyonic spin-0 states. One of them is the internal space volume mode, and the other is the lowest mode in the harmonic expansion of the dilaton. The rest of the complete spectrum is unitary. 
We also showed that once tiny perturbations in the two tachyonic modes are switched on, the dS$_4\times S^2$ vacuum flows to a solution asymptotic to Mink$_4\times S^2$ with minimal potential energy. The presence of the two tachyons can also be seen from the 4D effective action obtained below by substituting the ansatz for the metric, U(1) field strength and dilaton of the form 
\bea
d\hat{s}^2_6&=&e^{\frac12(\sigma+\varphi)}ds^2_4+
e^{-\frac12(\sigma+\varphi)}(d\theta^2+\sin^2\theta d\phi^2)\ ,
\nn\\
\hat{F}_{(2)}&=&\frac{p}2\sin\theta d\theta\wedge d\phi,\quad \hat{\phi}=\frac12(\varphi-\sigma)\ ,
\label{ra1}
\eea
and setting to zero all the other fields, which leads to standard Einstein Hilbert term and canonically normalized scalar kinetic terms with the potential
\bea
e^{-1}{\cal L}_{4D} &=& \frac14 \Big[ R-\frac12 (\partial_\mu \varphi)^2 -\frac12 (\partial_\mu\sigma)^2 -V_{4D}\Big]\ ,
\nn\w2
V_{4D}&=&\dfrac{e^{\varphi+\sigma}}{2(\tv e^{\sigma}+v e^{\varphi})}\left(4-4\tv e^{\s}+p^2\tv^2 e^{2\s}-4ve^{\varphi}+2p^2v\tv e^{\s+\varphi}+p^2v^2e^{2\varphi}\right)\Big]\ .
\label{eff}
\eea
This result is consistent in the sense that the 6D field equations are satisfied by keeping only the two scalars and the metric as in \eq{ra1}. For $p=1$ this becomes
\be
V_{4D}=\frac{e^{\varphi+\sigma}}{2(\tv e^{\sigma}+v e^{\varphi})}(2-\tv e^{\s}-ve^{\varphi})^2\ .
\ee
This potential, and its form for the special case of $\tv=0$, are displayed in Fig. \ref{fig:left} and Fig \ref{fig:right}.

The dS$_4$ and Mink$_4$ solutions have
\begin{align}
{\rm dS}_4:\quad&  e^{-\s}=3\tv,\quad e^{-\varphi}=3v,\quad V=\frac{4}{27v\tv}\ \Rightarrow \ L^2=\frac{81}2v\tv\ ,
\nn\\
{\rm Mink}_4:\quad & 2-\tv e^{\s}-ve^{\varphi}=0,\quad V=0\ .
\end{align}
The mass matrix around the dS$_4$ vacuum is 
\be
-\frac1{v\tv}\left( \begin{array}{cc} \frac{7}{54} &  -\frac{1}{54} \\
 -\frac{1}{54} &  \frac{7}{54}  \end{array} \right)
\ee
whose eigenvalues are $m_1^2=-6/L^2$ and $m_2^2=-9/(2L^2)$, in agreement with the spectrum analysis presented in the earlier sections. 

For completeness, let us also discuss the other case of $\b^2=3$ in which there is no flux on $S^2$. In this case, the corresponding effective potential from \eq{eff} with $p=0$ is given
\begin{align}
\b^2=3:\qquad V_{4D}=\frac{2e^{\varphi+\s}}{\tv e^\s+v e^\varphi}(1-\tv e^\s-v e^\varphi)\ .
\end{align}
This potential does not possess a lower bound. Instead, it exhibits a runaway behavior towards negative infinity. 

In our analysis, we have neglected contributions from higher order in derivative terms, as well as terms that arise from their supersymmetrization, which can contribute to the lower derivative terms as well as the potential in presence of gauging. However, a dimensional analysis in which the vacuum solution is considered suggests that their contribution will be suppressed for $v\tv\gg1$. To see this, let us consider the Lagrangian $\cL= \cL_2 +\Delta \cL$ with $\cL_2$ from \eq{Lagrangian} and $\Delta \cL$ expected to contain terms nonvanishing for our solutions, schematically taking the form:
\begin{align}
\Delta \cL &= \frac{1}{\k^2} \Big[ \kappa (a^{+} e^{\phi}+ a^{-}e^{-\phi}) \left( {\rm Riem}^2  + F^2\right) 
 +\k^3 (a^{+}\tv+a^{-} v) (a^{+} e^\phi+a^{-} e^{-\phi} ) f F^4 +\cdots \Big]\ ,
 \end{align}
where we have used the notation $a^\alpha= (a^{+}, a^{-})$ to avoid confusion with the $S^2$ radius $a$. The full 4-derivative extension of the R-symmetry gauged model with GS counterterm has not yet been constructed. Nonetheless the  terms in $\Delta\cL$ are motivated as follows. The term proportional to ${\rm Riem}^2$ has been discussed in \cite{Duff:1996rs,Bossard:2024ffp}, albeit in the context of ungauged theory. The $F^2$ term is unusual. At the tree level, it is a consequence of $U(1)_R$ gauging in presence of the ${\rm Riem}^2$ term, as was discovered in \cite{Bergshoeff:2012ax}. We have introduced an expected 1-loop dependent partner to it proportional to $a^{-} F^2$ here. The $fF^4$ term is expected to arise in presence of mixed gravitational and gauge anomaly, at least in the absence of $U(1)_R$ gauging, as was noted in \cite{Bossard:2024ffp}. We also expect corrections to the potential, possibly of the form $\sqrt{-g}/f^2$. However, such a term, unlike the lower order term displayed in \eq{pot} does not lend itself to an infinite expansion in $e^{-2\phi}$. 
Evaluated on the the dS$_4\times S^2$ solution, in which we recall that $e^{\phi}= \sqrt{\tv/v}$, and taking into account the quantization condition $L^2/a^2=9/2$, and using the relation $a^{+}=a^{-}=2$,  one can check that in the unit of $\kappa$
\begin{align}
& f F^2 \sim \frac1{\sqrt{v\tv}}\ , \quad f^{-1} \sim \frac1{\sqrt{v\tv}}\ ,\quad 
(a^{+}e^{\phi}+ a^{-} e^{-\phi})\, {\rm Riem}^2 \sim \frac1{\sqrt{v\tv} }\left(\frac1{v}+\frac1{\tv}\right)\ ,
\nn\w2
& (a^{+}\tv+a^{-} v) (a^{+} e^\phi+a^{-}e^{-\phi} ) f F^4 \sim \left(\frac1{v}+\frac{1}{\tv}\right)^2\ .
\end{align}
Thus the terms in $\Delta\cL$ are suppressed relative to the terms in the 2-derivative action for $v\gg1$ and $\tv\gg1$. In this regime, the $S^2$ radius $a\gg \sqrt{\k}$. If were to include a term of the form $\sqrt{-g}/f^2$ in the potential, it would also be suppressed in a similarly compared to the lowest order term in the potential.

For the two models described in Sec. 2.2, we have $v\tv$ around 1 or $10^{-2}$ and therefore they do not satisfy these conditions. 
One way to increase $v,\, \tv$ is to assign large U(1)$_{R^+}$ charges to hyperini. Doing that, we have found several models in which the desired conditions $v\gg 1$ and $v\gg 1$ are satisfied. These are tabulated in Appendix A. In the case of Mink$_4\times S^2$, and using $a^{+}=a^{-}=2$,  the suppression of the 4-derivative terms does not necessarily require $v$ and $\tv$ to be very large, but rather $a^2\gg\kappa \cosh\phi_0$, and $a^2\gg (v+\tv) \kappa$.

The model we have adopted is weakly coupled  when coupling constant  $\langle e^{-2\phi}\rangle$ is small.  For Mink$_4\times S^2$ and dS$_4\times S^2$ vacua, the weak coupling condition amounts to $\phi_0\gg1$ or equivalently $v/\tv \ll 1$, respectively. For $U(1)_{R+}$ and $U(1)_{R+}\times U(1)$ gauged models described in the text, and the $U(1)_{R+}$ models tabulated in Appendix C, the value of $v/\tv \approx 0.1$. Thus the weak coupling condition is marginally satisfied by all these models. If, on the other hand, $\langle e^{-2\phi}\rangle$ is large, the proper description should be in terms of the dual theory in which the coupling constant is  $\langle e^{2\phi}\rangle$. Hence, the model is genuinely strongly coupled only when $\tv/v\sim 1$, where a weakly coupled description is infeasible.

 In this paper, we have also considered dS$_6$ and dS$_2 \times S^4$ solutions.  The spectrum around dS$_6$ is summarized in Table 6, and it has two tachyonic scalars \footnote{It would be a significant progress to find a (meta)stable $6D$ de Sitter space solution in which the spectrum satisfies the FL bound \cite{Montero:2020rpl, Montero:2021otb}.}. 
In the case of dS$_2\times S^4$, the Maxwell flux on dS$_2$ is turned on, and consequently there is no need for flux quantization. In this case, the effective Lagrangian in $2D$ is a generalized version of the JT model, and we showed that the would-be tachyonic dilaton is absent for a particular value of the $U(1)_{R+}$ flux for which gives a scalar in the discrete unitary representation with conformal dimension $\Delta=3$\footnote{It is in interesting open problem to determine if this solution can be obtained as an extremal limit of a black hole solution.} This unitarity is due to the presence of a rigid shift symmetry at the linearized level, absent in the Lagrangian with interactions. 

We have also found nonsupersymmetric (A)dS$_4\times S^2$ solutions, when we turn on a flux associated with external $U(1)$ gauge field only, and show that they support the phenomenon of scale separation under certain conditions on anomaly coefficients. The one model with $U(1)_{R+}\times U(1)'$ symmetry we have have constructed does not satisfy the condition we have determined. However, it is very intriguing to search for solutions that satisfy the desired property in a wider class of models. 

The remarkable consequences of the GS counterterm for the vacuum structure of the theory, in particular yielding dS$_4$ solution, motivates a detailed computation of the one-loop effective action to tame the fuller implications of the quantum corrections. 
In particular, it remains to be investigated if the KK mass spectrum we have found will undergo a renormalization.
A systematic search for anomaly free models with diagonal gauging is also well motivated. The AdS involving vacua, of which we have given an example, remain to be investigated. Noting that the theory admits dS$_6$ vacuum also motivates search for black hole solutions with that asymptotics, to explore further the consequences of the GS term \footnote{It would be interesting to investigate possibility of viewing the dS$_4\times S^2$ solution as certain magnetic version of Nariai solution, arising as an extremal limit of a black three-brane solution in $6D$.}. These questions are relatively more manageable versions of similar ones one could pose for the heterotic string theory effective action, where the supersymmetrization of the GS counterterms are much more complicated 8-derivative terms, as opposed to the 4-derivative ones in $6D$, in which case R-symmetry gauging brings in new challenges not present in $10D$. 

\subsubsection*{Acknowledgments}

The authors thank Guillaume Bossard, Axel Kleinschmidt, Henning Samtleben, Ryo Suzuki, and Yuji Tachikawa for useful discussions.  The work of Y. P. is supported by the National Key R\&D Program No. 2022YFE0134300 and the National Natural Science Foundation of China (NSFC) under Grant No. 12175164, No. 12247103. The work of E. S. is supported in part by the NSF grant PHY-2413006 and TUBITAK, and X.G. is supported in part by George P. and Cynthia Mitchell Institute for Fundamental Physics and Astronomy. E.S. would like to thank the Department of Physics at the Istanbul Technical University for hospitality during the course of this work. E.S.’s visit to Istanbul Technical University was supported by TUBITAK “2221-Fellowships for Visiting Scientists and Scientists on Sabbatical Leave” with number 1059B212500261.

\begin{appendix}

\section{\texorpdfstring{$U(1)_{R+}$}{U(1)R+} gauged models with \texorpdfstring{$v\gg1$}{v>>1} and \texorpdfstring{$\tv \gg1$}{Tilde v >>1}}

In Section 7 we saw that the higher derivative corrections to our solutions are suppressed for $v\gg1$ and $\tv\gg1$ which are related to the anomaly coefficients $(b^+_{R+}, b^-_{R+})$ as in \eq{vb}. Here we give more $U(1)_{R+}$ models with large $v,\,\tv$ which we have obtained by dividing the  245 hyperini into 2 groups, in which $n$ of them carry charge $q$ and the rest carry charge $p$. Due to the time limitation, we only present a few examples up to $n=2$, and we have imposed an upper bound on the charge, i.e. $p,\,q\le 10000$.  Note that most of these examples are weakly coupled, as the effective coupling $\langle e^{-2\phi}\rangle\propto b^+_{R+}/b^-_{R+}\sim 0.1$.

\begin{table}[H]
    \centering
    \begin{tabular}{|c|c|c|c|c|c|}
    \hline
    $n $  & $q$ & $p$ & $b^-_{R+}$ &  $b^+_{R+}$ & $ b^+_{R+}/b^-_{R+}$\\ \hline
    1 & 185 & 101 & 186542 &23732 & 0.13 \\ \hline
    1  &185 & 115 &241502 & 30260 & 0.13\\ \hline
    2  &1 & 21 & 7940 & 992 & 0.12\\ \hline
     2  &1 & 23 & 9524 & 1190 & 0.12\\ \hline
      2  &1 & 25 & 11252 & 1406 & 0.12\\ \hline
     2  &1 & 27 & 13124 & 1640 & 0.12\\ \hline
      2  &1 & 29 &15140 & 1892 & 0.12\\ \hline
     2  &1 & 31 &17300 & 2162 & 0.12\\ \hline
      2  &1 & 33 &19604 & 2450 & 0.12\\ \hline
        2  &1 & 35 &22052 & 2756 & 0.12\\ \hline
       2  &1 & 37 &24644 & 3080 & 0.12\\ \hline  
        2  &1 & 39 &27380 & 3422 & 0.12\\ \hline 
          2  &1 & 9999 &1799640020 & 224955002 & 0.06\\ \hline 
    2  & 25 & 11 & 2232 & 324 & 0.15 \\ \hline
     2  &53 & 157 & 444212 & 55400 &0.12 \\ \hline
      2  &241 & 577 &6003570 & 747924 &0.12\\ \hline
       2  &1919 & 2419 &105980220 & 13127652 &0.12\\ \hline
    \end{tabular}
    \caption{In this table, we present more examples of $U(1)_{R+}$ gauged models with large anomaly coefficients $b^+_{R+}$ and $b^-_{R+}$ which are related to $v$ and $\tv$ as in \eqref{vb}.}
    \label{vector Love}
\end{table}

\section{The cases of dS\texorpdfstring{$_6$}{6} and dS\texorpdfstring{$_2 \times S^4$}{2 x S2}}

Here we consider the $U(1)_{R+}$ gauged model with no additional $U(1)$ symmetry.

\subsection{Spectrum around dS\texorpdfstring{$_6$}{6}}
\label{sec:ds6}

In this section we make a brief analysis of the linearized fluctuations around the maximally symmetric dS$_6$ vacuum. Our purpose is to identify the de Sitter representations of the elementary 6D fields and determine if there are any tachyonic states.  The background solution is given by
\begin{equation}
R_{\mu\nu}=\frac{5}{L^2}g_{\mu\nu}=\frac{1}{4\sqrt{v\tilde v}} g_{\mu\nu},
\quad
H_{(3)} = 0,
\quad
F_{(2)} = 0,
\quad
\phi_0=\frac12\ln\left(\frac{\tv}{v}\right)
\label{dS6background}
\end{equation}
Note that the solution requires that $v\tv>0$. Denoting the fluctuations around this background by $(\delta g_{\mu\nu}, \delta \phi, \delta A_\mu, \delta B_{\mu\nu},\delta \psi_\mu, \delta\chi,\d\l) := (h_{\mu\nu}, \varphi, a_\mu, b_{\mu\nu},\psi_\m,\chi,\l)$, we have the linearized field equations 
\begin{align}
& \left(\Box_6 - \frac{2}{L^2}\right)
h_{\mu\nu}+\frac{2}{L^2}g_{\m\n}h=0\ ,\quad
\left(\Box_6 - \frac{5}{L^2}\right)a_\mu=0 \ ,\quad
\left(\Box_6 - \frac{8}{L^2}\right)b_{\mu\nu}=0\ ,
\nn\w2
&\left(\Box_6 + \frac{10}{L^2}\right)\varphi = 0 \ ,\quad
\G^{\m\n\r}\nabla_\n\Tilde{\psi}_\r=0 \ ,\quad
\G^\m\nabla_\m\chi=0 \ ,\quad
\G^\m\Tilde{\psi}_\m=0\ ,\quad
\label{6deom}
\end{align}
where $h:=g^{\mu\nu}h_{\mu\nu}$, and we have defined the supersymmetric combination $\Tilde{\psi}_\m:=\psi_\m+2if\nabla_\m\l$, which represents the gravitino field that is massive as a result of eating the gaugino associated with $U(1)_{R+}$. We choose the gauges
\begin{align}
\nabla^\m h_{\m\n}=\frac12\nabla_\n h \ ,\qquad
\nabla^\m a_\m=0\ ,\qquad
\nabla^\m b_{\m\n}=0\ .
\end{align}
Next, we expand the fluctuation fields in terms of the Wigner functions associated with the coset $SO(1,6)/SO(1,5)$, and the $SO(1,5)$ irreps labeled by the a triplet of integers $(m,n,k)$ \footnote{For bosons, these integers corresponding to the number of boxes in each row of the Young tableaux.}, and the $SO(1,6)$ irreps by $(\Delta ,a,b)$, where $\Delta$ is the conformal weight. Denoting the Wigner functions by $D^{(\D ab)}_{(mnk)}$, the standard formula that computes the difference between the second order Casimir eigenvalues of $SO(1,6)$ and $SO(1,5)$ gives
\be
\Box_6 D^{(\D ab)}_{(mnk)}=-\frac{1}{L^2}\left[\D(\D-5)+a(a+3)+b(b+1)-m(m+4)-n(n+2)-k^2\right] D^{(\D ab)}_{(mnk)}\ .
\ee
\begin{table}
\centering
\begin{tabular}{|c|c|c|}
\hline
 $SO(1,6)$ content $(\D, a,b)$ & $SO(1,5)$ content $(m,n,k)$ & \mbox{Comments}\\
\hline\hline
$(5,2,0)$  & $(2,0,0)$ & $h_{\m\n}$ massless\\
\hline
$(\frac52,\frac32,\pm\frac12)$  & $(\frac32,\frac12,\pm\frac12)$ & $\Tilde{\psi}_\m$ massive
\\
\hline
$(4,1,0)$& $(1,0,0)$ & $a_\m$ massless
\\
$(3,1,1)$& $(1,1,0)$ & $b_{\m\n}$ massless
\\
\hline
$(\frac52, \frac12, \pm\frac12)$&  $(\frac12,\frac12,\pm\frac12)$ & $\chi$ massless
\\
\hline
$(\frac12+\frac12\sqrt{65},0,0)$ & $(0,0,0)$ & $\phi$ tachyon
\\
$(\frac12+\frac12\sqrt{65},0,0)$ & $(0,0,0)$ & $h$ tachyon
\\
\hline
\end{tabular}
\caption{The spectrum around dS$_6$ consist of one massless spin-2 field, one massless vector field, one massless two-form\cite{Basile:2016aen}, there are two scalars which are tachyonic states. In the fermionic sector, there is one massive spin-3/2 field which eats the gaugino \cite{Letsios:2023nonunitarity}, one massless spin-1/2 field.}
\end{table}
In Table 1, we have listed the $SO(1,5)$ and $SO(1,6)$ representation content associated with the fluctuations which obey the field  equations \eq{6deom}. Given the values of the conformal dimensions listed in this table, and using the criteria provided in \cite{Basile:2016aen}, we see that the fluctuations $(h_{\mu\nu}, a_\mu, b_{\mu\nu},\chi)$ are in the exceptional series, $\Tilde{\psi}_\m$ is in the principal series, and $(\varphi,h)$ are non-unitary representations. Thus, the $dS_6$ vacuum exhibits an instability.

\subsection{Comments on the case of dS\texorpdfstring{$_2\times S^4$}{2 x S2}}
\label{sec:ds2}

Here we shall consider a $dS_2\times S^4$ solution, which is non-supersymmetric and given by
\begin{align}
R_{\m\n}=\frac{1}{L^2}g_{\m\n}\ ,\ R_{ij}=\frac{3}{a^2}g_{ij}\ ,\ F_{\m\n}=k\epsilon_{\m\n}\ .
\end{align}
where $\m,\n=0,1$ and $i,j=2,3,4,5$, and $\epsilon, L, a, k, \phi$ are given by
\begin{align}
&\ \phi=\frac12\ln\left(\frac{\tv}{v}\right)\ ,\qquad \frac{1}{L^2}=-3k^2\sqrt{v\tv}+\frac{1}{4\sqrt{v\tv}}\ ,\qquad \frac{1}{a^2}=\frac13 k^2\sqrt{v\tv}+\frac{1}{12\sqrt{v\tv}}\ .
\label{ds2sol}
\end{align}
Since $dS_2$ is non-compact, there is no flux quantization condition. It also follows from the solution that 
\begin{align}
dS_2 \times S^4 : & \qquad 0\leq k^2<\frac{1}{12v\tv}\ .
\end{align}
At $k^2=\dfrac{1}{12v\tilde v}$, one obtains Mink$_2 \times S^4$, while for $k^2>\frac{1}{12v\tilde v}$ the solution becomes Ad$S_2 \times S^4$. Note also that in the case of $dS_2 \times S^4 $ the flux parameter $k$ is allowed to vanish. 
To study the stability condition for this vacuum, we start with the ansatz 
\bea
d\hat{s}^2_6&=&e^{-3\sigma}ds^2_2+
e^{2\sigma}d\Omega_4^2\ ,
\quad \hat{\phi}=\phi\ ,
\quad \hat{A}_\m=A_\m\ ,
\eea
where $d\Omega_4^2$ is the metric on round $S^4$, and we obtain the 2D effective action
\begin{align}
\mathcal{L}_2=\frac14\sqrt{-g}\left(e^{4\s}R+12e^{-\s}-e^{7\s} fF_{\m\n}F^{\m\n}-\frac{2}{f}e^\s-e^{4\s}(\partial_\m \phi)^2\right)\ ,
\label{JT}
\end{align}
where $f=v e^\phi+\tv e^{-\phi}$. The resulting EOMs are
\begin{align}
\d\phi:&\quad \nabla^\m\left(e^{4\s}\partial_\m\phi\right)-e^{7\s} f' F_{\m\n}F^{\m\n}+\frac{2f'}{f^2}e^\s=0\ , 
\label{2deom1}
\\
\d\s:&\quad4e^{4\s}R-12e^{-\s}-7e^{7\s} f F_{\m\n}F^{\m\n}-\frac{2}{f}e^\s-4e^{4\s}\partial_\m\phi\partial^\m\phi=0\ ,
\label{2deom2}
\\
\d{g^{\m\n}}:&\quad\left(g_{\m\n}\Box_2-\nabla_\m\nabla_\n\right)e^{4\s}+\left(-6e^{-\s}+\frac12e^{7\s} f F^2+\frac1f e^\s+\frac12 e^{4\s}\partial_\r\phi\partial^\r \phi\right)g_{\m\n}
\nonumber\\
&-2e^{7\s} f F^2_{\m\n}-e^{4\s}\partial_\m\phi\partial_\n\phi=0\ ,
\label{2deom3}
\\
\d{A_\m}:&\quad \nabla^\m\left(e^{7\s} f F_{\m\n}\right)=0\ .
\label{2deom4}
\end{align}
Focusing on the solutions that do not have spacetime dependence, we take the ansatz
\begin{align}
\phi={\rm constant}\ ,\qquad \sigma={\rm constant}\ ,\qquad F_{\m\n}=ke^{-3\sigma}\epsilon_{\m\n}\ .
\end{align}
Thus, \eqref{2deom4} is satisfied, and \eqref{2deom1}-\eqref{2deom3} become 
\begin{align}
2f' e^\s \left(k^2+\frac{1}{f^2}\right)&=0\ , \label{2d1}
\\
2e^{3\s}R-6e^{-2\s}+7k^2f-\frac1f&=0\ ,\label{2d2}
\\
-6e^{-\s}+e^\s k^2f+e^\s \frac1f&=0\ .\label{2d3}
\end{align}
Next, \eqref{2d1} is solved by $f'=0$ which gives $e^{2\phi}=\tv/v$. This, in turn, implies that $f=2\sqrt{v\tv}$. In addition, $\s$ can be solved from \eqref{2d3}:
\begin{align}
e^{-2\s}=\frac16\left(k^2f+\frac1f\right)\ ,\label{s4r}
\end{align}
which agrees with \eqref{ds2sol}. The curvature of $dS_2$ is fixed by \eqref{2d2}:
\begin{align}
e^{3\s}R=-3k^2f+\frac1f\ ,\label{ds2r}
\end{align}
which also agrees with \eqref{ds2sol}. The positivity of $dS_2$ curvature requires that
\begin{align}
3k^2f<\frac1f\quad \implies \quad 0\leq k^2<\frac{1}{12v\tv}\ .
\end{align}
Thus, we reproduce the $dS_2\times S^4$ solution \eqref{ds2sol}.

\subsubsection*{Stability of the dS$_2 \times S^4$ solution}

We saw that the 2D Lagrangian \eq{JT}  admits the dS$_2$ vacuum solution with
\be
R=e^{-3\sigma}\left(\frac{1}{f}-3k^2 f\right)\ ,\qquad 
e^{-2\s}=\frac16\left(k^2f+\frac1f\right)\ ,\qquad e^{2\phi} =\frac{\tv}{v}\ ,\qquad F_{\m\n}=ke^{-3\sigma}\epsilon_{\m\n}\ .
\ee
The Lagrangian \eq{JT} has the form of Jackiw-Teitelboim (JT) gravity generalized to include couplings to the fields $\phi$ and $A_\mu$. Examining the fluctuations around the about the above solution, for reasons well understood in the analysis of the standard JT gravity, here too one can show that the metric and  the scalar $\sigma$ are non-dynamical. We also see from \eq{2deom4} that the vector field does not propagate either. Thus we are left with \eq{2deom1} for the dilaton. Denoting the dilaton fluctuation by $\delta\phi$, it follows that
\be
\Box_2 \d \phi +12e^{-5\s} \d \phi=0\ .\label{dt}
\ee
It follows that $\delta\phi$ has a negative mass squared 
\be
m^2=-12e^{-5\s} \ .
\ee
Denoting $l$ to be the radius $dS_2$ from \eqref{ds2r} 
\begin{align}
    \frac{1}{l^2}=\frac12e^{-3\s}\left(\frac1f-3k^2f\right)=\frac{1}{L^2}e^{-3\s}\ ,
\end{align}
we can read from \eqref{dt}
\begin{align}
\left(-\frac{\D(\D-1)}{l^2}+12e^{-5\s}\right)\d\phi=0\ ,
\end{align}
by using \eqref{s4r}, if follows that
\begin{align}
\D=\frac12\pm\frac12\sqrt{1+16\frac{4k^2v\tv+1}{1-12k^2v\tv}}\ .
\end{align}
Since $k^2<1/(12v\tv)$, $\d\phi$ cannot sit in the principal series or in the complementary series. In general $\d\phi$ is non-unitary, however, when $\D\in\mathbb{Z}_+$, $\d\phi$ can be unitary and sits in the discrete series\cite{Sun:2021thf}. For example, when $k^2=1/(20v\tv)$, we have $\D=3$. If we consider the vacuum solution for $k=0$, the conformal dimension of $\delta$ is fixed to belong to non-unitary representation.

The discrete representations of de Sitter group in 2D are discussed in detail in \cite{Farnsworth:2024dS2}\cite{Anninos:2023DiscreteDS2}. These are unitary representation which are special due to the fact that the field equation that describes them has a rigid shift symmetry. For example, in the case of $\Delta=3$ we found above, this symmetry takes the form $\delta\phi = S_{AB} X^A X^B$, where $A=0,1,2$, and $X^A$ is the standard embedding of the d$S_2$ into 3D flat Lorentzian space by the relation $\eta_{AB}X^AX^B=1/l^2$, where 
$\eta={\rm diag}(-1,1,1)$, and $S_{AB}$ is a constant tensor which is symmetric and traceless\cite{Farnsworth:2024dS2}. Let us note however that this symmetry does not hold for the interacting Lagrangian \eq{JT}. 

\section{ (A)dS\texorpdfstring{$_4\times S^2$}{4 x S2} solutions with \texorpdfstring{$U(1)'$}{U(1)'} flux and scale separation}
\label{sec:ads}

In this appendix we consider the $U(1)_{R+}\times U(1)'$ model with only the $U(1)'$ flux turned on. We consider the ansatz
\begin{align}
&R_{\mu\nu}=\frac{3\e}{L^2}g_{\mu\nu}\ ,\qquad
R_{ij}=\frac{1}{a^2}g_{ij}\ , \qquad F'_{ij}=c\epsilon_{ij}\ , \qquad \mbox{rest}=0\ ,
\end{align}
where
\begin{align}
\e=
\begin{cases}
-1\ , & \text{for AdS$_4\times S^2$}\ ,\\
+1\ , & \text{for dS$_4\times S^2$}\ .
\end{cases}
\end{align}
The Maxwell and the 2-form equations of motion as well as the Bianchi identities are satisfied automatically, and the remaining equations are
\begin{align}
\mbox{Dilaton}:\qquad &c^2=\frac{f'}{f^2}\frac{1}{f'_1}\ ,
\label{ddp}
\\
\mbox{Einstein} \ (\mu\nu):\qquad& \frac{3\e}{L^2}=-\frac12f_1c^2+\frac{1}{2f}\ ,\label{e1p}
\\
\mbox{Einstein}\ (ij):\qquad&\frac{1}{a^2}=\frac32f_1c^2+\frac{1}{2f}\ ,
\label{e2p}
\\
\mbox{Quantization condition:} \qquad &\ ca^2 =\frac{n}{2}\ ,\quad n\in \mathbb{Z}\ ,
\label{qc2p}
\end{align}
where $f=ve^\phi_{R+} + \tv_{R+} e^{-\phi}$ and $f_1:=v_1 e^\phi+\tv_1 e^{-\phi}$. Defining
\begin{align}
\lambda := \frac{f_1 f'}{f_1' f}\ ,
\end{align}
we obtain
\begin{align}
\frac{1}{L^2} = \frac{\e(\lambda-1)}{6f}\ ,
\quad
\frac{1}{a^2} = \frac{3\lambda+1}{2f}\ ,
\quad
c^2 = \frac{\lambda}{ff_1}\ ,\quad 
\frac{f}{f_1} = \frac{n^2(3\lambda+1)^2}{16\lambda}\ .
\label{nep}
\end{align}
Solving the last equation in \eqref{nep} gives
\begin{align}
e^{2\phi}=\frac{n^2 \tv_1 (3 \lambda+1)^2-16 \tv \lambda}{16 v \lambda-n^2 v_1 (3\lambda+1)^2}\ .
\label{e2phip}
\end{align}
Substituting this into the second equation in \eq{nep} yields
\begin{align}
&81 n^4 v_1 \tv_1 x^4+ 72\e x^3 \left(6 n^4 v_1 \tv_1- n^2 v \tv_1- n^2 \tv v_1\right)+
\nonumber
\\
&x^2 \left(864 n^4 v_1 \tv_1-336 n^2 v \tv_1-336 n^2 \tv v_1\right)+256\e x \big(3 n^4 v_1 \tv_1-2 n^2 v \tv_1
\nn\\
& -2 n^2 \tv v_1+2 v \tv\big)+ 256(v-n^2 v_1) ( \tv-n^2 \tv_1)=0\ ,
\label{qep}
\end{align}
where
\begin{align}
x=\e(\lambda-1)\ .
\end{align}
The existence of a vacuum solution depends on whether this quartic equation has at least one positive real root. A sufficient condition is
\begin{align}
v_1 \tv_1\left(v-n^2 v_1\right) \left(\tv-n^2 \tv_1\right)<0\ .
\label{sufficientn}
\end{align}
It also follows from \eq{nep} that the existence of the solutions require that $\tv$ and/or $tv_1$ must be nonvanishing.  For the model given in \eq{m2}, the anomaly coefficients are given
\be
(v,\tv)= \frac{1}{\sqrt{32\pi^3}}(38,12)\ ,\qquad (v_1,\tv_1)= \frac{1}{\sqrt{32\pi^3}} (15,98)\ ,
\ee
hence the inequality \eqref{sufficientn} is satisfied only for $n=\pm1$ in this case.
Indeed, \eq{qep} has the real solution $x=1.419$ for d$S_4\times S^2$ and $x=2.542$ for Ad$S_4\times S^2$. We also note that if the condition
\be
\left(v-n^2 v_1\right) \left(\tv-n^2 \tv_1\right) \approx 0\ ,
\ee
holds, then \eq{qep} admits a root $x\approx 0$, for which scale separation occurs since
$L^2/a^2= 3(4+3x)/x \gg 1$ for $x \approx 0$. Note that the model we have given in \eq{m2}, this condition is not satisfied. However, we expect that in a wider class of models with exterior abelian fields activated, the desired conditions may be satisfied.

\section{Group theoretical formulae}

\subsection{Unitary representations of de Sitter group and Higuchi bound in 4D}

\subsubsection*{\boxed{\it Bosons}}

For scalar fields obeying the equation
\be
(\Box-m^2)\phi=0 \ \implies \ m^2=\frac{\Delta(3-\Delta)}{L^2}\ .
\label{sc1}
\ee
In $4D$ de Sitter space, the unitary representations arise as follows:\cite{Sun:2021thf}
\begin{align}
\mbox{Principal series:} \qquad & m^2 > \frac{9}{4L^2} \ \implies \ \Delta=\frac32 +i\mu\ , \  \mu\in \mathbb{R}\backslash\{0\}\ ,
\nn\w2
\mbox{Complementary series:} \qquad & 0 < m^2 \le \frac{9}{4L^2} \ \implies \ 0<\Delta <3\ ,
\nn\w2
\mbox{Exceptional series:}\qquad &  \Delta=0,-1,-2,\dots
\label{hb}
\end{align}
In the last equation the value of $m^2$ is as defined in \eq{sc1}. Turning to massive spin-s fields for $s\ge 1$ in dS$_4$, they  have the field equations \cite{Bianchi:2005ze}
\be
(\Box-m^2_{dS}-m^2)h_{\mu_1\cdots\mu_s}=0\ ,\quad\nabla^{\m_1}h_{\m_1\cdots\m_s}=0,\quad g^{\m_1\m_2}h_{\m_1\m_2\cdots\m_s}=0\ ,
\label{A3}
\ee
where 
\be
m^2_{dS}=\frac{s(2-s)+2}{L^2}\ .
\label{dSmassless}
\ee
In 4D de Sitter space, we have
\begin{align} 
\Box_4 h_{\mu_1...\mu_s} 
= -\frac{\Delta(\Delta-3)-s}{L^2}h_{\mu_1...\mu_s}\ .
\end{align}
It follows that
\be
m^2=-\frac{\Delta(\Delta-3)-(s-2)(s+1)}{L^2}\ .
\label{nm}
\ee

According to \cite{Sun:2021thf}, the unitary representation labeled by $(\Delta,s)$ of $SO(1,4)$ are as follows:
\begin{align}
&{\cal F}_{\Delta,s} \quad,\quad\Delta=\frac{3}{2}+i\mu \quad,\quad\left(\mu\in \mathbb{R}\backslash\{0\},s\in \mathbb{Z}_+ \right) \quad\mbox{Principal Series}
\label{h1}\\
&{\cal F}_{\Delta,s} \quad,\quad1<\Delta<2 \quad,\quad(s\in \mathbb{Z}_+) \quad\mbox{Complementary Series}
\label{h2}\\
&{\cal F}_{\Delta,s} \quad,\quad\Delta=2+t \quad,\quad(t=0,1,\dots, s-1,s\in \mathbb{Z}_+) \quad\mbox{Exceptional Series}
\label{h3}
\end{align}
In particular, $t=s-1$ describes a strictly de Sitter massless, i.e. i$m^2=0$, while $t=0,1,...,s-2$ describe {\it partially massless} spin-s states. In terms of $m^2$ defined in \eq{nm}, the unitarity conditions are
\begin{align}
\mbox{Principal series:} \qquad & m^2 \ge \frac{1}{L^2} \left(s-\frac12\right)^2\ ,
\nn\w2
\mbox{Complementary series:} \qquad & \frac{1}{L^2} s(s-1) < m^2 < \frac{1}{L^2} \left(s-\frac12\right)^2\ ,
\nn\w2
\mbox{Exceptional series:}\qquad & m^2=\frac{(s-t-1)(s+t)}{L^2}\ .
\label{hb1}
\end{align}
The Higuchi bound is the lower bound for complementary series. 

\subsubsection*{\boxed{\it Fermions}}

Fermion fields with arbitrary half integer spin $s:=r+\frac12$, $r \in \mathbb{Z}_+$ and mass parameter $m$ on $dS_4$ can be described by Dirac gamma-traceless and divergent-free totally symmetric tensor-spinors $\Psi_{\mu_1...\mu_r}$ obeying the field equation \cite{Deser:2003gw,Deser_2001,Letsios:2023nonunitarity}:
\begin{align}
(\slashed{\nabla}  -m)\psi_{\mu_1...\mu_r}=0\ ,
\ \nabla^\nu\psi_{\nu\mu_2...\mu_r}=0\ ,\ 
\gamma^\nu\psi_{\nu\mu_2...\mu_r}=0\ ,
\label{A13}
\end{align}
Expanding the tensor spinor as in \eq{fhe}, and using the formula
\begin{align}
\slashed{\nabla}D^{(\Delta,s)}_{(\mu_1...\mu_r)\pm}
=\pm \frac{1}{L}\left(-\Delta+\frac32\right)D^{(\Delta,s)}_{(\mu_1...\mu_r)\mp}\ ,
\label{ds1}
\end{align}
we get
\be
\psi_+\left(\frac32-\Delta\right)=Lm \psi_-\ ,\quad \psi_-\left(\frac32-\Delta\right)=-Lm\psi_+ ,
\ee
where we have suppressed the labels $(\Delta,s)$. For this equation to have nontrivial solution, we find
\be
\left(\frac32-\Delta\right)^2=-(Lm)^2\ .
\ee
According to \cite{Letsios:2023nonunitarity}, the unitary representation labeled by $(\Delta,s)$ of $SO(1,4)$ are as follows:
\begin{align}
&{\cal F}_{\Delta,s} \quad,\quad\Delta=\frac{3}{2}+i\mu \quad,\quad \left( \mu\in \mathbb{R}^+\backslash\{0\}\right)  \quad\mbox{Principal Series}
\label{h1f}\\
&{\cal F}_{\Delta,s} \quad,\quad\Delta= 3/2\quad,\quad  \qquad\mbox{Discrete Series}
\label{h2f}\\
&{\cal F}_{\Delta,s} \quad,\quad\Delta=2+t \quad,\quad\left(t=\frac12,\dots, s-1;\ s\in \mathbb{Z}_+-\frac12 \right) \quad\mbox{Discrete Series}
\label{h3f}
\end{align}
In particular, for $s\geq\frac32$, $t=s-1$ describes a strictly de Sitter massless, while $t=\frac12,...,s-2$ describe {\it partially massless} spin-s states. For $s=\frac12$, $\D=\frac32$ describes a massless de Sitter massless state.

\subsection{Properties of harmonics on \texorpdfstring{$S^4$}{S4} and dS\texorpdfstring{$_4$}{4}}

The formulae below hold for $S^4$. To get the corresponding ones for dS$_4$, we send $n\to -\Delta$, and to obtain the formula for AdS$_4$, we send $\nabla\to i\nabla$ and $n\to -E_0$ \cite{Sezgin:1984ku}. 

\begin{align}
\Box_4 D^{(n0)}(x) &= -\frac{1}{L^2}n(n+3) D^{(n0)}(x)\ ,
\\
\Box_4 D^{(nn_1)}_\mu (x) & = -\frac{1}{L^2}[n(n+3)+n_1(n_1+1)-3]D^{(nn_1)}_\mu (x)\ , 
\\
\Box_4 D^{(nn_1)}_{(\mu\nu)}(x) &= -\frac{1}{L^2}[n(n+3)+n_1(n_1+1)-8]D^{(nn_1)}_{(\mu\nu)}(x)\ , 
\label{g3}\\
\Box_4 D^{(n1)}_{[\mu\nu]\pm}(x) &= -\frac{1}{L^2}[n(n+3)-2]D^{(n1)}_{[\mu\nu]\pm}(x)\ ,
\\
\nabla_\mu D^{(n0)}(x) &=\frac{1}{2L}\sqrt{n(n+3)}D^{(n0)}_\mu (x)\ ,
\\
\nabla^\mu D^{(n1)}_\mu (x) &=0\ , 
\\
\nabla^\mu D^{(n0)}_\mu (x) &=\frac{2}{L}\sqrt{n(n+3)}D^{(n0)}(x)\ ,
\\
\nabla^\mu D^{(n2)}_{(\mu\nu)} & = 0\ ,
\\
\nabla^\mu D^{(n1)}_{(\mu\nu)} & =\frac{3}{2L}\sqrt{(n-1)(n+4)}D^{(n1)}_\nu\ ,
\\
\nabla^\mu D^{(n0)}_{(\mu\nu)} & = \frac{3}{2L}
\sqrt{\frac{3}{2}(n-1)(n+4)}D^{(n)}_\nu\ ,
\\
\nabla^\mu D^{(n1)}_{[\mu\nu]\pm} & = \pm \frac{1}{2L}\sqrt{3(n+1)(n+2)}D^{(n1)}_\nu\ ,
\\
\nabla_\mu D^{(n0)}_\nu & = \nabla_\nu D^{(n0)}_\mu\ ,
\\
\nabla_\mu D^{(n1)}_\nu & - \nabla_\nu D^{(n1)}_\mu = 
\frac{1}{L}\sqrt{\frac{1}{3}(n+1)(n+2)}(D^{(n1)}_{[\mu\nu]+} - D^{(n1)}_{[\mu\nu]-})\ .
\end{align}
In the fermionic sector, denoting the flat indices on $S^4$ by $a=0,1,2,3$, we have 
\begin{align}
    \gamma^\mu\nabla_\mu D^{(n,\frac12)}_\pm &= \pm \frac{1}{L}\left(n+\frac{3}{2}\right) D^{(n,\frac12)}_{\mp}\ , \quad
    \\
    \gamma^\mu\nabla_\mu D^{(n,\frac12)}_{a\pm} &= \pm \frac{2}{5L}\left(n+\frac32\right) D^{(n,\frac{1}{2})}_{a\mp}\ , \quad
    \\
    \gamma^\mu\nabla_\mu D^{(n,\frac32)}_{a\pm} &= \pm \frac{1}{L}\left(n+\frac32\right) D^{(n,\frac32)}_{a\mp}\ , \quad
    \label{A3/2}
    \\
    \nabla^a D^{(n,\frac{3}{2})}_{a\pm} &= 0 \ , \quad
    \\
    \nabla^a D^{(n,\frac12)}_{a\pm} &= \mp \frac{3}{2L}\sqrt{\left(n-\frac12\right) \left(n+\frac72\right)} D^{(n,\frac12)}_{a\mp}\ , \quad
\end{align}

\end{appendix}

\newpage

\providecommand{\href}[2]{#2}\begingroup\raggedright


\end{document}